\pdfoutput=1

\documentclass[10pt,conference]{IEEEtran}
\IEEEoverridecommandlockouts

\usepackage{caption}
\usepackage{subcaption}
\usepackage{graphicx}
\usepackage{adjustbox}
\usepackage{hyperref}
\usepackage{algorithm}
\usepackage{algpseudocode}
\usepackage{amsmath,amsfonts}
\usepackage{booktabs}
\usepackage{flushend}

\newcommand{\nsf}[1]{\href{https://www.nsf.gov/awardsearch/showAward?AWD_ID=#1}{#1}}

\author{\IEEEauthorblockN{Anthony Etim}
\IEEEauthorblockA{
\textit{Yale University}\\
New Haven, CT, USA\\
anthony.etim@yale.edu}
\and
\IEEEauthorblockN{Jakub Szefer}
\IEEEauthorblockA{
\textit{Northwestern University}\\
Evanston, IL, USA\\
jakub.szefer@northwestern.edu}
}

\begin{document}

\title{Fault Attacks on ML-based Quantum Control\\ and Error Correction}

\pagestyle{plain}

\maketitle

\begin{abstract}
    Machine-learning (ML) models are increasingly used in quantum computing systems to discriminate multi-qubit readouts, mitigate correlated readout errors, and decode quantum error-correcting codes, making them an integral component of today's quantum computer control and readout stacks. This paper is the first to analyze the susceptibility of such ML models to physical fault injection, which can cause quantum computers to return incorrect results or perform wrong error correction operation. This work studies two representative architectures: (i)~a fully connected neural network for 5-qubit (32-class) readout error correction (HERQULES), and (ii)~a convolutional neural network used as a Deep~Q-learning (Deep Q) decoder for the distance-5 Surface Code. Using the ChipWhisperer Husky for voltage glitching together with automated search over the fault parameter space, this work localizes successful fault settings to specific layers of each target ML model. On the HERQULES model, fault susceptibility is strongly layer-dependent: early layers exhibit higher misprediction rates than later layers. On the Deep~Q decoder, a single trigger-aligned voltage glitch in either the first convolutional layer or the final fully connected output layer reduces decoding accuracy from $100\%$ to as low as $21.57\%$. We further characterize the resulting failures at the bitstring level using Hamming-distance and per-bit flip statistics, showing that single-shot glitches can induce structured corruption rather than purely random noise. These results motivate treating ML-based quantum readout and error-correction decoding as security-critical components, and highlight the need for lightweight fault-detection and redundancy mechanisms in quantum computing~pipelines.

\end{abstract}


\section{Introduction}
\label{sec_introduction}

Quantum computers rely on measurement of analog signals from qubit readout logic to convert qubit states into classical bitstrings that can be processed by classical computers. While cross-talk errors among qubits and gates have been steadily decreasing due to improvements in quantum computer hardware, the readout errors and noise remain some of the dominant error sources in near-term quantum systems. To mitigate these errors, quantum readout logic increasingly deploys ML-based readout pipelines: from neural discriminators for multiplexed readout to hardware-efficient ML architectures that scale with qubit counts~\cite{lienhard2022dnnreadout,maurya2023herqules}. Beyond readout discrimination, ML models are also being deployed as decoders for quantum error-correcting codes, where reinforcement-learning-trained convolutional neural networks (CNNs) can decode surface codes in fault-tolerant settings~\cite{sweke2021rl_decoders}. While these approaches improve fidelity and enable more robust quantum computation, they also introduce a new security-critical dependency, which is the ML algorithm responsible for performing readout correction or error-correction decoding~operations.

\begin{figure}[t]
    \centering
\includegraphics[width=0.9\linewidth]{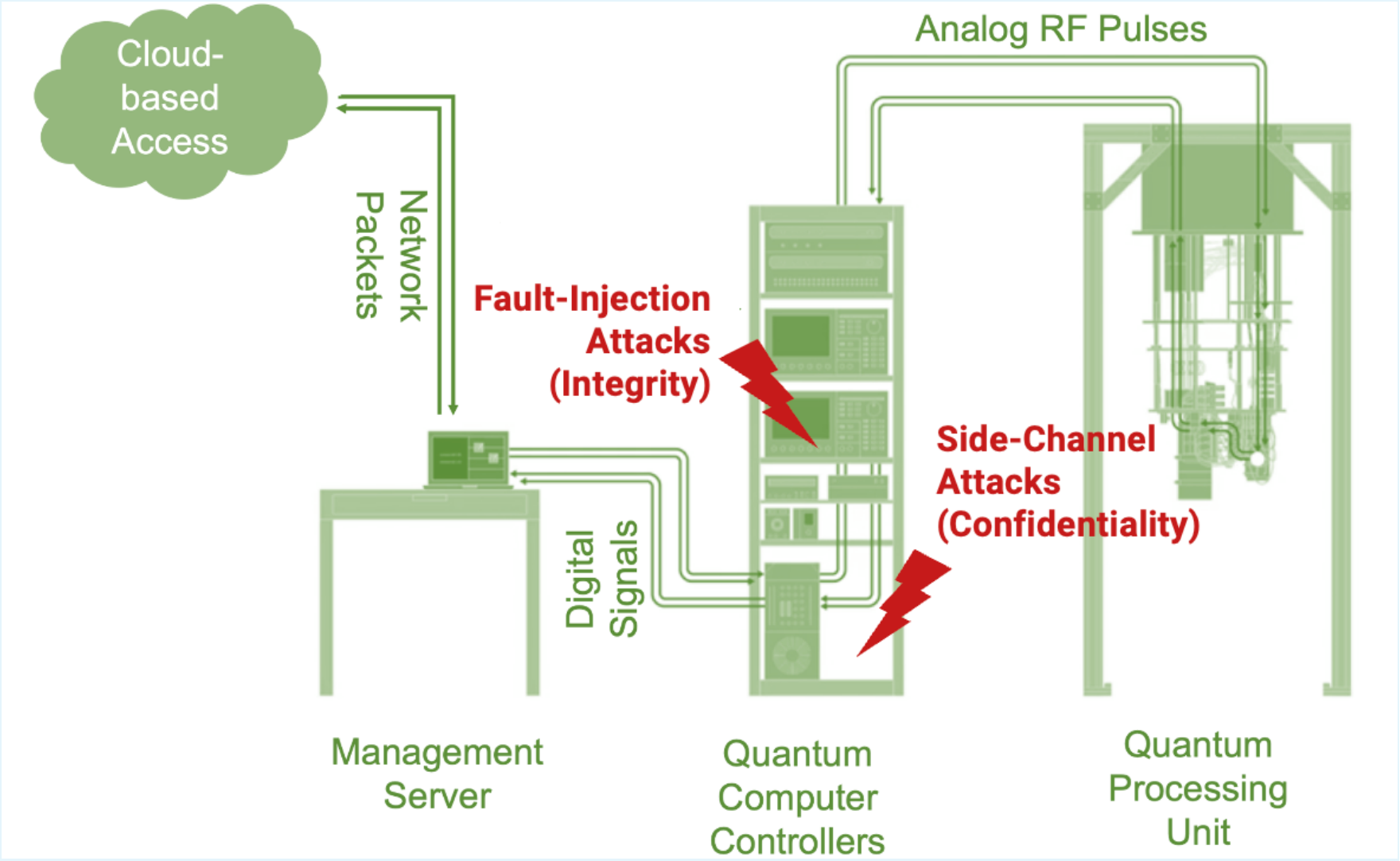}
    \caption{Quantum computer stack and the two attack surfaces it exposes. Fault-injection attacks (this work) target the classical controllers; side-channel attacks (out of scope) target the analog signals between controllers and the Quantum Processing Unit (QPU).}
    \label{fig:threats}
\end{figure}

Recent works on the security of the quantum computers highlight that cloud deployment and untrusted classical infrastructure create new threats around the quantum processors~\cite{trochatos2024dynamic,xu2024quantum,xu2023classification,ghosh2023primer}.
Among existing security analysis of quantum computers, readout logic and associated controllers and algorithms are not well understood. A diagram of a modern quantum computing system with fault injection vulnerabilities in the controller is shown in Figure~\ref{fig:threats}.

To address this research gap, in this paper we focus on one concrete and under-studied area: \emph{physical fault injection} against ML models used in quantum computer readout error correction and error-correction decoding. We study two representative ML architectures: (i)~HERQULES~\cite{maurya2023herqules}, a fully connected neural network for multi-qubit readout error correction, and (ii)~a Deep~Q-learning (Deep Q) decoder~\cite{sweke2021rl_decoders} based on a CNN for decoding the distance-5 surface code. 

Through physical experimentation, in this work we show that a voltage-glitch adversary with physical access to the classical controller can induce mispredictions, structured biases in HERQULES outputs and a collapse of Deep~Q decoder outputs onto a single dominant action. We use voltage glitching because it needs only commodity equipment and brief access to the controller's supply rail, requires no decapsulation or persistent device modification, and leaves no on-device artifact, properties that make it a standard first-choice fault primitive over more invasive (e.g., laser) or more easily detected (e.g., power interruption) alternatives, as we detail in Section~\ref{sec:threat_model}. The results are critical as these outputs are necessary to extract information from the quantum computer and to maintain logical qubit integrity, and cannot be interpreted if the readout or decoding logic produces wrong outputs. Beyond untargeted corruption, the induced failures are structured rather than random: a single glitch can drive the Deep~Q decoder to one fixed action regardless of the input syndrome and can bias HERQULES toward specific readout classes (Section~\ref{sec:results}). This constitutes a limited but concrete form of output control; producing an arbitrary attacker-chosen output on arbitrary inputs is harder and is left to future work. These physical fault injections can be performed without any modification to the quantum computer as they target the classical controllers and embedded ML algorithms, not the qubits that may be hard to access or manipulate.

\noindent\textbf{Contributions.} We present the first hardware fault-injection study of embedded ML models used in quantum computing, targeting both a readout error-correction model and a quantum error-correction decoder. Our contributions are:

\begin{itemize}
    \item We formalize a practical physical voltage-glitch adversary against embedded ML-based quantum readout error correction and quantum error-correction decoding in the quantum--classical control path.
    \item We study two representative ML architectures: (i)~HERQULES, a fully connected network for 5-qubit readout error correction, and (ii)~a Deep~Q-learning CNN decoder for the distance-5 surface code, analyzing fault susceptibility across both dense and convolutional layers.
    \item We implement a repeatable, trigger-synchronized ChipWhisperer Husky voltage-glitch workflow that localizes injections to specific neural network~layers.
    \item We use Optuna~\cite{akiba2019optuna} for automated search in HERQULES and Deep Q-learning over glitch parameters to efficiently find high-impact fault settings under constraints.
    
    \item We quantify layer-wise fault susceptibility across both models, comparing architecturally distinct layer types (dense vs.\ ReLU vs.\ output; convolutional vs.\ fully connected) and characterizing how specific readout and decoding outputs degrade at each layer.
\end{itemize}

 \section{Background}
\label{sec:background}
\label{background}

This section provides information on quantum computer systems' readout logic as well as fault injection attacks on machine learning algorithms.

\subsection{Quantum Measurement and Readout}

Quantum computers rely on a classical measurement interface that maps a quantum state to discrete outcomes. This readout is noisy due to instrument drift, amplifier noise, crosstalk, and state-preparation-and-measurement (SPAM) effects~\cite{nachman2020unfolding}, and the errors can be \emph{biased} and \emph{correlated} across qubits, which is especially consequential for multi-qubit syndrome measurements~\cite{rojkov2022bias}. Even when gate errors are reduced, measurement error can remain a dominant contributor to logical failure unless explicitly mitigated~\cite{maciejewski2020readout,nation2021measurement_mitigation,ghosh2023primer}.

\subsection{ML-based Readout Error Correction}

ML is increasingly leveraged in the readout path. Deep networks capture non-linear boundaries and correlations from multiplexed readout, crosstalk, and non-Gaussian noise, improving discrimination and calibration robustness~\cite{lienhard2022dnnreadout,kim2021deeplearning_readout}; HERQULES in particular emphasizes \emph{hardware-efficient} post-processing suitable for embedded, low-latency execution~\cite{maurya2023herqules}. Because such models sit on the critical path of syndrome extraction and control, their integrity is tied to the trustworthiness of the whole computation~\cite{xu2024quantum,trochatos2024dynamic}.

\subsection{Quantum Error Correction}
Quantum Error Correction (QEC) encodes one logical qubit in many physical qubits; stabilizer measurements yield \emph{syndromes} indicating likely errors without collapsing the logical state. A classical \emph{decoder} maps the (noisy) syndrome history to corrective Pauli operations, and decoder errors translate directly into logical errors~\cite{xu2024quantum,trochatos2024dynamic}. Classical decoders such as minimum-weight perfect matching (MWPM) are strong baselines for simple noise, but ML decoders are attractive because they learn correlated, non-Pauli noise, tolerate syndrome measurement errors, and amortize inference into one forward pass~\cite{sweke2021rl_decoders,fitzek2020deepq_toric,bausch2024alphaqubit}.

\subsection{ML-based Decoders for Quantum Error Correction}

Beyond readout discrimination, ML models are also used as decoders for quantum error-correcting codes such as the surface code~\cite{sweke2021rl_decoders,andreasson2019toric_rl,fitzek2020deepq_toric,bausch2024alphaqubit,lange2023gnn_decoder}. In the reinforcement-learning formulation, the decoder observes the syndrome history as a state~$s$, selects a Pauli correction~$a$, and is trained to maximize a reward given only on successful logical recovery; the policy is a Q-function $Q(s,a)$ and the decoder picks $\arg\max_a Q(s,\cdot)$. \emph{Deep} Q-learning parameterizes $Q$ with a neural network, scaling to surface-code state spaces. Sweke et~al.~\cite{sweke2021rl_decoders} realized this for noisy syndromes with a CNN whose input is a stack of $(2d{+}1)\times(2d{+}1)$ binary matrices (faulty syndrome slices and action histories) and whose output is Q-values over corrective actions.

\subsection{Fault Injection Attacks}

\begin{figure}[t]
\centering
\includegraphics[width=0.9\linewidth]{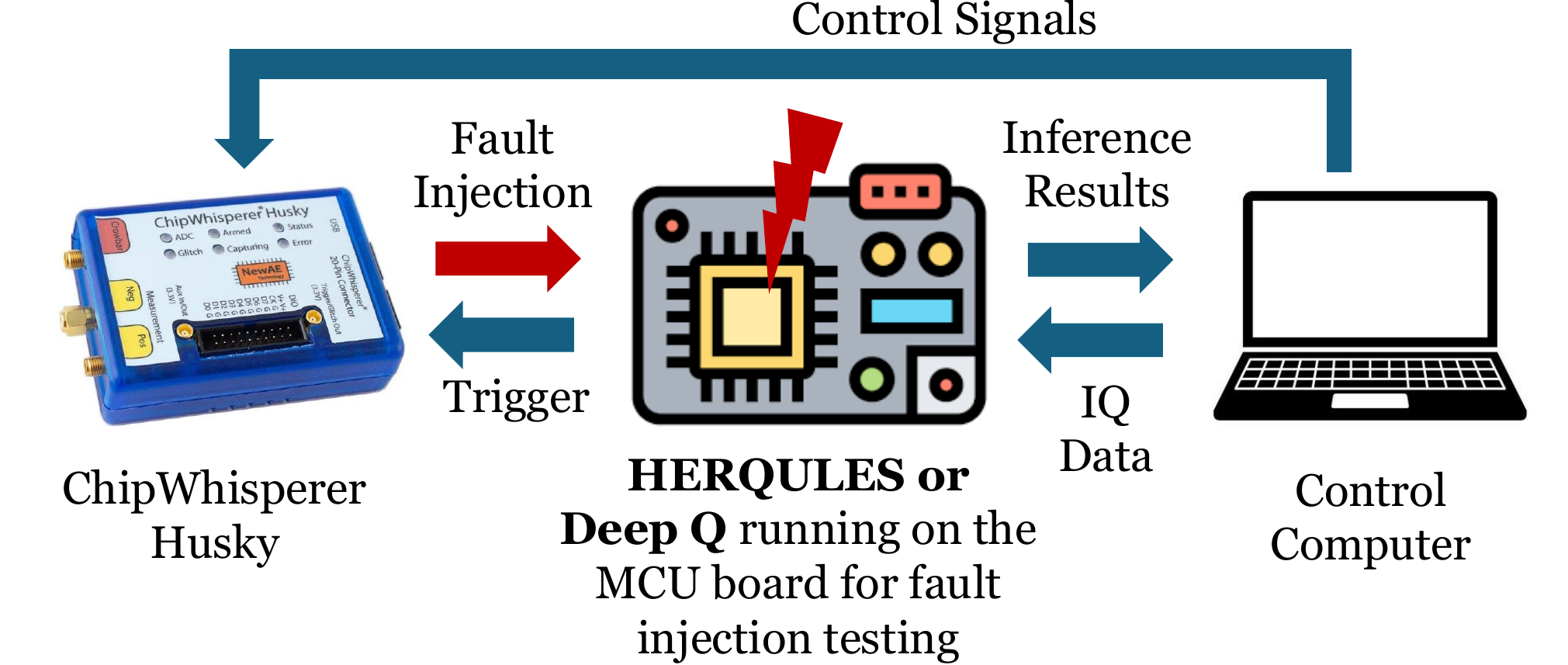}
\caption{\small  Overview and setup for evaluation of physical fault injection against ML models that perform quantum readout error correction and quantum error-correction (QEC) decoding. The host PC orchestrates inference queries and logging. ChipWhisperer Husky injects voltage glitches on the target's supply rail, aligned to a trigger emitted by the target at the start of a chosen neural network layer. Each trial returns a predicted class (or decoder action) and a status (correct, misprediction, or reset or hang)}
\label{fig_workflow}
\end{figure}

Fault injection in classical systems is known to perturb device operation, e.g., via voltage or clock glitching or electromagnetic pulses, and is able to induce transient computation errors~\cite{barenghi2012fi_survey,eslami2020survey_fi,gangolli2022iot_fi_review,erata2024systematic}. Further, on ML workloads, faults may corrupt activations, intermediate accumulators, or control flow, potentially causing misclassification, targeted output manipulation, or silent integrity violations~\cite{breier2018faultdnn}. These lessons motivate our evaluation of ML-based readout correction  under fault attacks and lead us to {\em find new fault injection attacks on quantum computer readout and error correction.}

\section{Threat Model}
\label{sec:threat_model}
\label{sec:threatmodel}

\subsection{Adversary's Capabilities}

We assume an attacker with physical access to the quantum computer controller where the ML algorithm inference is running (e.g., malicious insider, compromised maintenance workers, or other attackers with momentary access), enabling transient fault injection through voltage glitching. We stress that this controller is a \emph{classical} computer, the embedded processor or SoC on which the readout-correction and decoder inference executes, not the quantum processor itself, so physical access to it is access to a conventional classical device rather than to the qubits or any cryogenic hardware. The attacker can: (i) physically glitch the voltage supply of the controller to induce voltage glitches in the ML algorithm as it executes on the controller, and
(ii) synchronize a trigger signal to the execution of the ML inference to cause the short-duration voltage glitches to occur in specific parts of the ML~execution. 

We assume the attacker has ability to control the glitches parameterized by: {\tt width} or duration of the glitch, intra-cycle {\tt offset} of the glitch, {\tt external offset} within the target program (here, the ML algorithm) that determines when the glitch is injected, and {\tt repeat} number of glitches that should be applied. These options are common in platforms such as ChipWhisperer devices~\cite{oflynn2014chipwhisperer}. The attacks do not require persistent modification of firmware or stored weights of the ML models. Further, there is no modification to the quantum computing dilution refrigerator or any hardware running at cryogenic temperatures. All attacks are done at the controller, outside of the dilution refrigerator making them much easier than any attacks on components inside the dilution~refrigerator.

\subsection{Adversary's Goals}
\label{adversary_goals}
The attacker's primary objective is \emph{integrity violation} of the ML model's outputs, whether those outputs are corrected readout bitstrings (HERQULES) or corrective Pauli actions selected by a surface-code decoder (Deep~Q).
We consider two goals:
\begin{itemize}
    \item \textbf{Untargeted Degradation:} reduce the model's accuracy on its intended task. For readout correction, this means increasing the Hamming distance between the predicted and true bitstrings, leading to incorrect interpretation of the quantum computation's results. For QEC decoding, this means producing corrective actions that disagree with the syndrome-conditioned ground truth, leading to logical errors that compound across decoding rounds.
    \item \textbf{Targeted Steering:} bias outputs toward attacker-chosen values. For readout correction, this means biasing corrected bitstrings toward specific classes; for QEC decoding, it means steering the decoder toward a specific corrective action (or a small set of actions) regardless of the input syndrome, which causes the decoder to apply a fixed recovery rather than the syndrome-conditioned one the surface code's fault-tolerance argument assumes.
\end{itemize}

\subsection{Attacker Motivations and Scenarios}
\label{sec:motivation}

We focus on malicious insiders or other adversaries inside the data center who have incentives to sabotage operation of the quantum computer, or to covertly drive readout or QEC results to a target result.
Within that context, several concrete scenarios make ML readout and decoding inference an attractive attack surface:

\begin{itemize}
     \item \textbf{Maintenance Access:} Quantum control electronics are physically large, frequently serviced, and sit in racks accessed by maintenance staff and third-party integrators; a compromised technician or a brief unattended window suffices to instrument a controller's supply rail, with no cryogenic access or quantum-hardware modification~needed.

     \item \textbf{Calibration Sabotage:} Modern superconducting devices rely on calibration loops that feed readout statistics back into pulse and frequency tuning. A persistent induced bias in the corrected readout, such as the structured ``$00001$'' bias we observe under HERQULES Layer-2 faults, silently corrupts calibration and causes long-lived fidelity degradation that outlives the attack.

\item \textbf{QEC Decoder Neutralization:} For fault-tolerant systems the ML decoder is the sole component between physical-qubit and logical errors; intermittently driving it to return the same action regardless of syndrome (which a single glitch achieves) breaks the syndrome-conditioned correction while still emitting well-formed outputs that no obvious check would flag.

\end{itemize}

\noindent 
Physical fault injection is also preferred over simpler alternatives. \emph{Cutting power} or denying service produces an immediately visible failure the provider can detect and recover from, and \emph{modifying firmware or stored weights} leaves persistent artifacts (changed hashes, flash writes, audit-log entries) usable to reconstruct the attack. A trigger-aligned glitch leaves none of these traces, the device returns to its nominal state after every shot and only a statistical output bias remains, which is why glitching stays attractive even when more invasive alternatives exist~\cite{barenghi2012fi_survey,boneh1997faults}.

\section{Evaluation Setup}

We evaluate two representative ML components from the quantum control stack: a multi-qubit readout error correction model and a surface-code decoder. This section describes the two models and the embedded platform on which we run them. Our evaluation is organized around four research questions (RQs) that also determine the experiments and metrics in Section~\ref{sec:results}:
\begin{itemize}
  \item \textbf{RQ1 (Susceptibility):} Can a single trigger-aligned voltage glitch on the classical controller cause an embedded ML readout-correction model or QEC decoder to produce an incorrect output, without modifying firmware or stored weights?
  \item \textbf{RQ2 (Layer dependence):} How does fault susceptibility vary across layers and across architecturally distinct layer types (dense vs.\ ReLU vs.\ output for HERQULES; convolutional vs.\ fully connected for the Deep~Q decoder)?
  \item \textbf{RQ3 (Failure structure):} Are the induced failures diffuse degradation, or are they structured and attacker-favorable, steering outputs toward a fixed class or action rather than merely adding noise?
  \item \textbf{RQ4 (Mechanism and defense):} Is there an identifiable propagation mechanism that explains the dominant failure mode and that a low-cost defense could target?
\end{itemize}
We answer RQ1 and RQ2 through the layer-localized glitch campaigns, RQ3 through the bitstring-level and confusion-matrix analyses, and RQ4 through the non-finite propagation experiment, all in Section~\ref{sec:results}.

\subsection{Target ML Models}
\label{sec:target_models}
We consider two ML models deployed on classical controllers in the quantum computing stack:

\noindent\textbf{Target Model 1: HERQULES Readout Error Correction.}
We consider a superconducting quantum computing system in which raw qubit readout generates IQ samples (in-phase and quadrature analog samples of the dispersive readout signal), which are then processed by an ML model to output the correct classical values of the readout. We focus on a superconducting system where the readout logic is shared among qubits. In particular, $5$ qubits share one readout line~\cite{heinsoo2018rapid}, so the readout logic has to correct for any correlated errors in these bits. This setup is based on existing work which demonstrated an ML algorithm for error correction of the correlated errors~\cite{maurya2023herqules}. Given a $5$-bit input (from the $5$ qubits), there are $2^5=32$ classes for the ML algorithm to classify the input data into; output indicates the error-corrected readout values of the 5~qubits.

The classifier takes as input a $10$-dimensional vector of IQ samples (in-phase and quadrature components per qubit) and outputs a probability distribution over the $32$ candidate $5$-bit readout strings. The architecture consists of two fully connected hidden layers (Dense~1 and Dense~2), each followed by a ReLU activation (ReLU~1 and ReLU~2), and a final dense output layer that produces $32$ logits, one per candidate class. A softmax over those logits yields the class probabilities, and an argmax selects the predicted readout~\cite{maurya2023herqules}.

\noindent\textbf{Target Model 2: Deep Q-Learning (Deep Q) Surface Code Decoder.}
We additionally consider a distance-5 surface code whose syndrome decoding is performed by a Deep~Q-learning agent~\cite{sweke2021rl_decoders}. The decoder's Q-network is a CNN that takes as input a stack of $(2d{+}1) \times (2d{+}1) = 11 \times 11$ binary matrices encoding syndrome values and action histories, and outputs Q-values for corrective Pauli operations on each of the $d^2 = 25$ data qubits. 
The architecture consists of three convolutional layers (Conv1, Conv2, Conv3) with ReLU activations, followed by flattening and a fully connected hidden layer (FC1, $512$ units, ReLU + dropout). A final dense layer (FC2) produces $|\mathcal{A}| + 1$ outputs, of which one is the state value $V(s)$ and $|\mathcal{A}|$ are the per-action advantages $A(s,a)$; these are recombined into the Q-values $Q(s,a) = V(s) + (A(s,a) - \overline{A(s,\cdot)})$ as in the dueling-network configuration of~\cite{sweke2021rl_decoders}.
A fault in this decoder can cause incorrect corrective actions, leading to logical errors in the surface code.

 We focus on \emph{the classical controller and embedded processor} that runs these ML inferences, and their vulnerability to attacks, consistent with broader concerns that quantum workloads may run atop partially untrusted classical infrastructure~\cite{trochatos2024dynamic,xu2024quantum,xu2023classification}.

\subsection{Embedded Target and Transferability}
\label{sec:platform}
Quantum control stacks split work across hardware tiers: FPGA or RFSoC fabric handles low-latency RF generation and demodulation, while calibration, ML post-processing, decoder inference, and feedback orchestration typically run on colocated SoCs, microcontrollers, or CPUs~\cite{maurya2023herqules}. HERQULES and the Sweke et al.~Deep~Q decoder were both proposed in this software-on-embedded-target style. We ported both codes from Python to C, we then use a ChipWhisperer Husky with a CW312T-SAM4S microcontroller target (ATSAM4S2A,
Cortex-M4) for two reasons: cycle-accurate trigger synchronization (which is what makes the per-layer windows in Algorithms~\ref{alg:herqules_forward_cycles} and~\ref{alg:deepq_forward_cycles} possible), and reproducibility on commodity hardware.

The main limitation of this study is that a production low-latency controller may instead place these computations on FPGA or RFSoC fabric, which we do not validate directly. We focus on voltage glitching the code on an MCU, but the decoding and error correction can also run on FPGAs. Voltage glitching has a long track record against FPGA, RFSoC, and ASIC targets~\cite{barenghi2012fi_survey,eslami2020survey_fi,gangolli2022iot_fi_review}; the underlying primitive (driving the supply rail out of envelope so paths violate setup time or registers latch wrong values) does not care whether a layer is a software MAC loop or a hardware DSP slice. What changes is the temporal surface: a software MAC offers many cycles per activation, a pipelined MAC offers a narrow window per result but many results in parallel. Our finding that early, wide layers (HERQULES Dense~1, Deep~Q Conv1) fault more easily than narrow late layers fits this picture, so we expect ``early-and-wide is fragile'' to transfer, but we have not validated this on FPGA. Confirming the same outcomes on FPGA and RFSoC targets, and identifying new failure modes when convolution maps to a systolic array rather than a MAC loop, are left to future work.

\section{Attack Implementation}
\label{sec:attacksetup}

\begin{algorithm}[t]
\caption{\small HERQULES neural network forward pass with per-layer cycle windows (MCU cycles). Cycle counts correspond to our C implementation of the algorithm running on ChipWhisperer Husky.}
\label{alg:herqules_forward_cycles}
\small
\begin{algorithmic}[1]
\Require Input vector $x \in \mathbb{R}^{10}$ \Comment{IQ samples from the 5-qubit shared readout}
\Require Parameters $\{(W_1,b_1),(W_2,b_2),(W_o,b_o)\}$
\Ensure Output probabilities $\hat{y} \in \mathbb{R}^{32}$

\State \Comment{Layer 1: Dense 1 }
\State $z_1 \gets W_1\, x + b_1$

\State \Comment{Layer 2: ReLU 1 }
\State $a_1 \gets \mathrm{ReLU}(z_1)$ \Comment{$\mathrm{ReLU}(u) = \max(0, u)$, applied elementwise}

\State \Comment{Layer 3: Dense 2 }
\State $z_2 \gets W_2\, a_1 + b_2$

\State \Comment{Layer 4: ReLU 2 }
\State $a_2 \gets \mathrm{ReLU}(z_2)$

\State \Comment{Layer 5: Output layer}
\State $z_o \gets W_o\, a_2 + b_o$ \Comment{logits over 32 classes}
\State $\hat{y} \gets \mathrm{softmax}(z_o)$ \Comment{class probabilities}
\State \Return $\hat{y}$
\end{algorithmic}
\end{algorithm}

\subsection{Fault Injection Attack Workflow}
We port both ML models to C and execute them on a target Micro-Controller Unit (MCU) instrumented with ChipWhisperer Husky~\cite{oflynn2014chipwhisperer} to test the effect of voltage glitching on each algorithm. The HERQULES readout error correction model is ported from the reference implementation in~\cite{maurya2023herqules}, and the Deep~Q surface-code decoder is ported from the reference implementation in~\cite{sweke2021rl_decoders}; both run on the same MCU target shown in Figure~\ref{fig_workflow}, and both are attacked using the same workflow. The figure also summarizes the experimental loop. A host PC configures the ChipWhisperer Husky's scope and glitch module and communicates with the MCU target running the ML algorithm. For each trial, we reboot and arm the target to ensure a clean state, send one test input (IQ data from~\cite{maurya2023herqules} for HERQULES, or a syndrome volume from~\cite{sweke2021rl_decoders} for the Deep~Q decoder), inject a single glitch aligned to a trigger, and log the returned output (a 5-bit class for HERQULES, a corrective action index for Deep~Q) along with device status (correct, misprediction, or reset or hang).

\subsection{Faulting Parameter Search Space}
In each voltage fault experiment we explore four parameters: \textbf{Glitch Width} (\texttt{width}), the duration of a single glitch pulse expressed in phase-shift steps and swept over $[0, 4000]$; \textbf{Glitch Offset} (\texttt{offset}), the sub-cycle phase offset relative to the victim clock edge in phase-shift steps over $[0, 4000]$; \textbf{External Offset} (\texttt{ext\_offset}), the number of victim clock cycles from trigger to the first pulse, swept over $[0, E_{max}]$; and \textbf{Repeat} (\texttt{repeat}), the number of pulses per trigger over $[1, 5]$. The search space is explored using Optuna~\cite{akiba2019optuna}, given the fault parameters together with a model-specific metric of~success.

\begin{algorithm}[t]
\caption{\small Deep Q-learning decoder CNN forward pass. Layer-localized fault injection on Conv1 and FC2 uses trigger points emitted by the C implementation around each targeted layer.}
\label{alg:deepq_forward_cycles}
\small
\begin{algorithmic}[1]
\Require Syndrome volume input $S \in \mathbb{R}^{C_{\text{in}} \times 11 \times 11}$
\Require CNN parameters 
$\{(K_1,b_1^c),(K_2,b_2^c),(K_3,b_3^c),$
\Statex \hspace{\algorithmicindent}
$(W_1^f,b_1^f),(W_2^f,b_2^f)\}$
\Ensure Q-values $Q(s,a)$ for each action $a$

\State \Comment{Layer 1: Conv1}
\State $h_1 \gets \text{ReLU}(K_1 * S + b_1^c)$

\State \Comment{Layer 2: Conv2}
\State $h_2 \gets \text{ReLU}(K_2 * h_1 + b_2^c)$

\State \Comment{Layer 3: Conv3}
\State $h_3 \gets \text{ReLU}(K_3 * h_2 + b_3^c)$

\State $h_{\text{flat}} \gets \text{Flatten}(h_3)$ 

\State \Comment{Layer 4: FC1 (hidden, 512 units)}
\State $z_1 \gets \text{ReLU}(W_1^f h_{\text{flat}} + b_1^f)$

\State \Comment{Layer 5: FC2 (dueling output)}
\State $u \gets W_2^f z_1 + b_2^f$ \Comment{$u \in \mathbb{R}^{|\mathcal{A}|+1}$: value $V(s)$ and advantages $A(s,\cdot)$}
\State $Q(s,a) \gets V(s) + \bigl(A(s,a) - \overline{A(s,\cdot)}\bigr)$ \Comment{dueling combine}
\State \Return $Q$
\end{algorithmic}
\end{algorithm}

For HERQULES the objective maximizes the expected Hamming distance between the predicted 5-bit output and the ground truth, subject to a penalty for resets or hangs. For the Deep~Q decoder the objective maximizes the rate at which the decoder returns an action other than the syndrome-conditioned ground-truth action (equivalently, $1 - \text{accuracy}$), again subject to a reset or hang penalty. In both cases the metric captures \emph{untargeted} integrity violation; we discuss the targeted case in Section~\ref{adversary_goals} and leave a dedicated targeted-objective search to future work.

To limit the searching time, for HERQULES we randomly select test data corresponding to one of each of the 32 readout classes and fault each one 3 times, resulting in 96 fault attempts per tested parameter combination. For the Deep~Q decoder we sweep over the 26 (X-noise model) or 51 (depolarizing model) ground-truth syndrome-action pairs and similarly fault each one multiple times. Throughout this work, success of a fault parameter is measured as the number of faulty outcomes out of the total fault attempts at that parameter~combination.

\subsection{Synchronization and Per-Layer Timing Windows}
To attribute faults to specific neural network layers, we instrument trigger points around each layer of each model and empirically measure the cycle ranges where each layer executes. For HERQULES this means trigger points at the entry to \texttt{dense\_1()}, \texttt{relu\_1()}, and the remaining layers, and for the Deep~Q decoder it means trigger points at the entry to Conv1 and FC2 (the two architecturally distinct layer types we target). We then restrict the external-offset search to these layer-specific windows (Algorithms~\ref{alg:herqules_forward_cycles} and~\ref{alg:deepq_forward_cycles}), enabling controlled per-layer comparisons in both models.

\subsection{HERQULES: Model and Faulting Setup}
\label{sec:herqules_fault_setup}
As a first target, we port the HERQULES readout error-correction model from Maurya et~al.~\cite{maurya2023herqules} to C and deploy it on the ChipWhisperer Husky target MCU. Its two-dense-layer architecture is that of Target~Model~1 (Section~\ref{sec:target_models}); Algorithm~\ref{alg:herqules_forward_cycles} gives the per-layer cycle windows we use to localize glitches.

In our fault-injection on HERQULES, we target every layer to characterize how susceptibility varies along the inference~pipeline:
\begin{itemize}
    \item \textbf{Dense~1 and ReLU~1 (early layers):} These layers process the raw IQ input and execute the longest multiply-accumulate chains in the network. Their wide intermediate activations expose a large temporal surface to a glitch, and small numerical perturbations injected here can cascade through the rest of the pipeline before the~softmax.
    \item \textbf{Dense~2 and ReLU~2 (middle layers):} These layers operate on the already-reduced hidden representation. A fault here has fewer downstream non-linearities to traverse, but also a narrower input dimensionality to~corrupt.
    \item \textbf{Output layer (final dense + softmax):} This is the last layer before the argmax that determines the predicted class. A fault here can directly alter the predicted-class distribution, but the softmax normalization tends to absorb modest perturbations.
\end{itemize}

\subsection{Deep Q Decoder: Model and Faulting Setup}
\label{sec:deepq_setup}

As a second target, we port the Deep~Q-learning decoder from Sweke et~al.~\cite{sweke2021rl_decoders} to C and deploy it on the same ChipWhisperer Husky target MCU. Its CNN Q-network is that of Target~Model~2 (Section~\ref{sec:target_models}), with three convolutional layers, a $512$-unit fully connected layer (FC1), and a dueling output layer (FC2); Algorithm~\ref{alg:deepq_forward_cycles} summarizes the forward pass.

In our fault-injection campaign on the Deep~Q decoder, we target two layers that represent architecturally distinct computation patterns:
\begin{itemize}
    \item \textbf{Conv1 (first convolutional layer):} This layer applies learned kernels across the spatial syndrome input. Convolutional layers use shared weights (kernel parameters) across all spatial positions, so a fault corrupting a kernel weight or partial sum can propagate to many output feature map positions simultaneously.
    \item \textbf{FC2 (dueling output layer):} This is the layer that directly produces the value and advantage outputs from which the Q-values are formed; an argmax over those Q-values selects the corrective action. A fault here can directly alter which action the decoder selects without affecting earlier intermediate representations.
\end{itemize}

\section{Experimental Results}
\label{sec:results}
\label{experimental_results}

\subsection{Fault Injection Results for Attacks on ML-Assisted Readout}
\label{sec:herqules_results}

\subsubsection{Successful Faults Across Neural Network Layers}

\begin{figure*}[t]
    \centering
    \begin{subfigure}{0.32\textwidth}
        \centering
        \includegraphics[width=\linewidth]{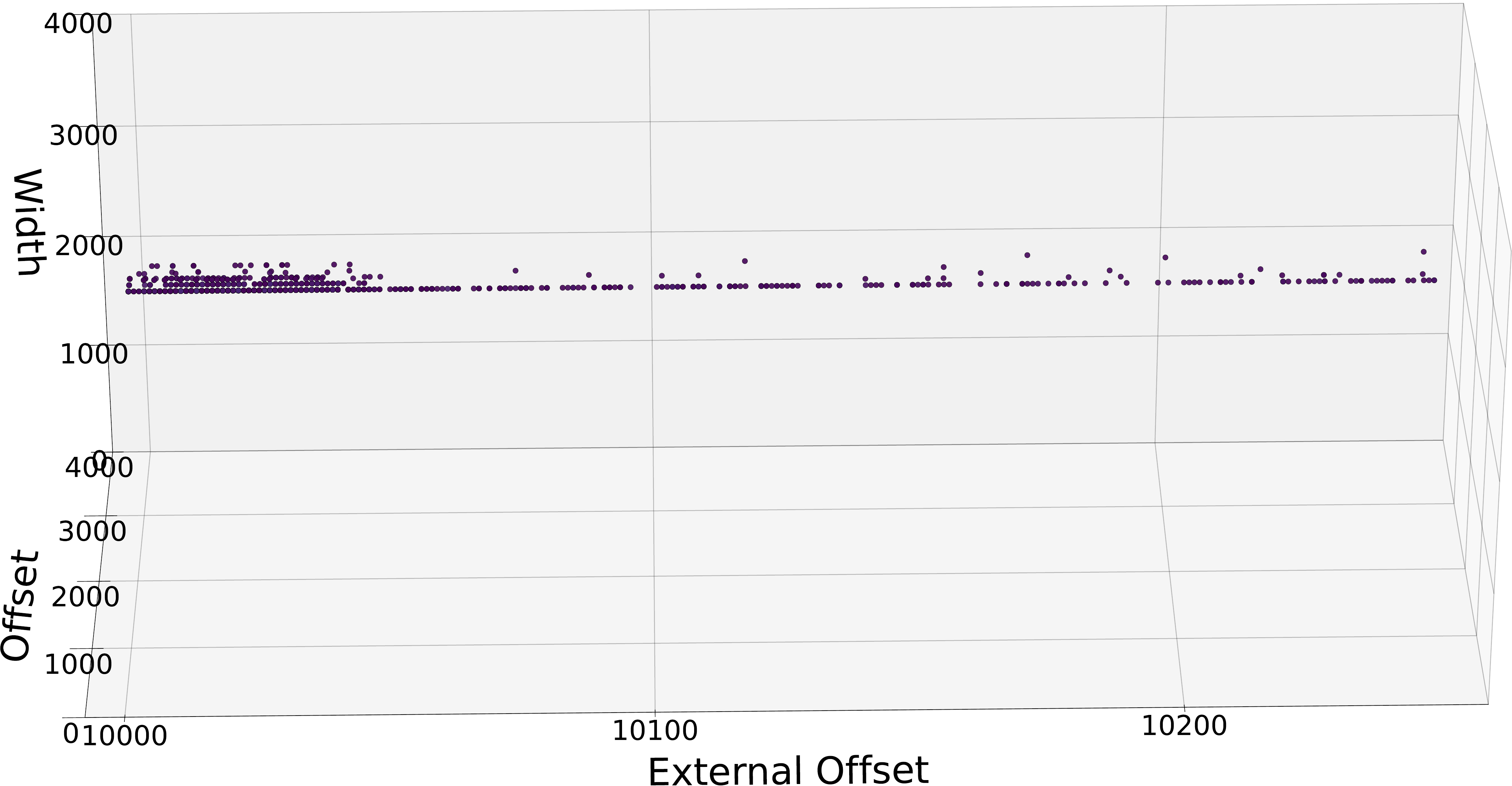}
        \caption{\small Layer 1: Dense 1}
        \label{fig:dense_1_3d}
    \end{subfigure}\hfil
    \begin{subfigure}{0.32\textwidth}
        \centering
        \includegraphics[width=\linewidth]{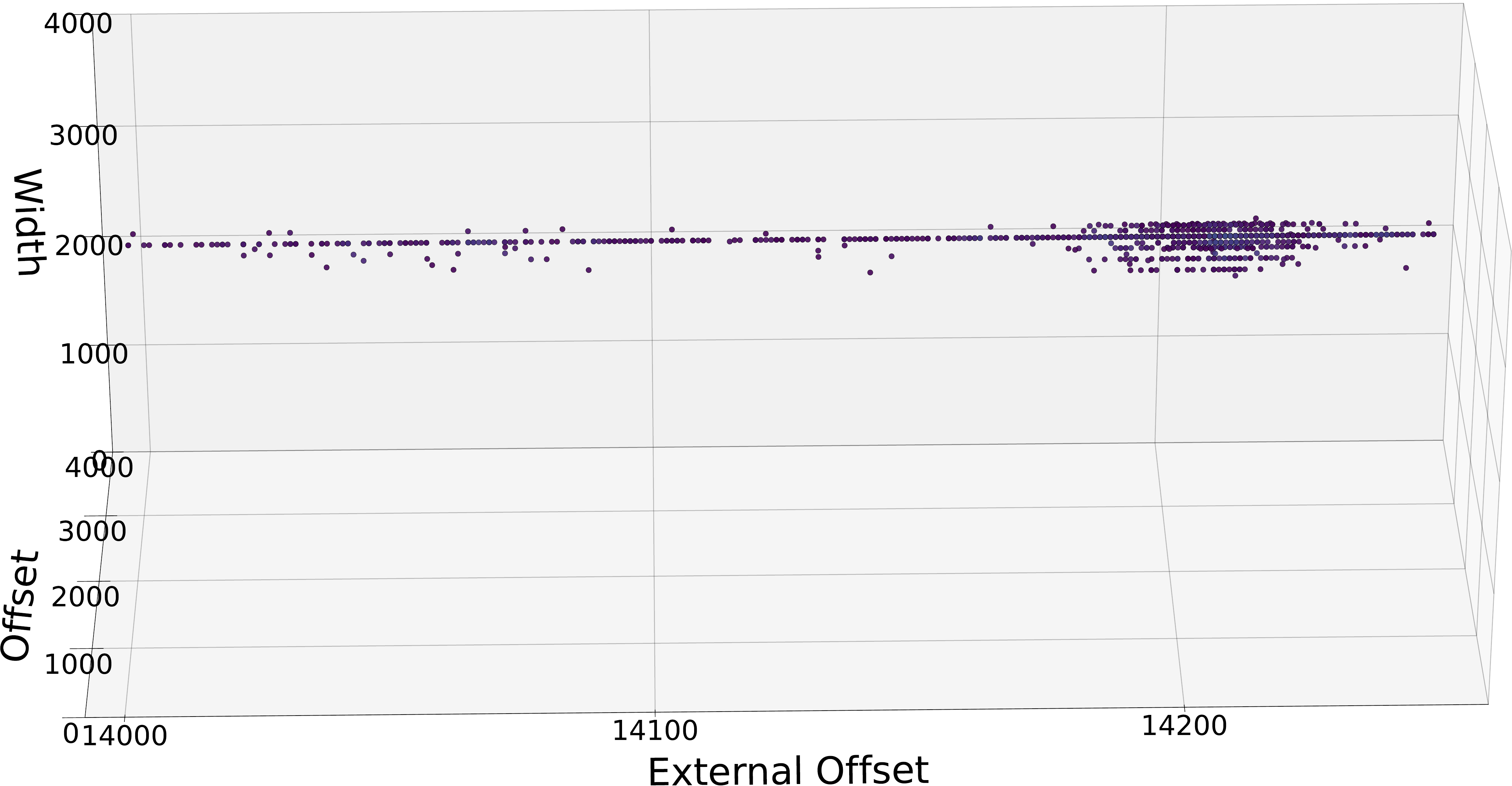}
        \caption{\small Layer 2: ReLU 1}
        \label{fig:relu_1_3d}
    \end{subfigure}\hfil
    \begin{subfigure}{0.32\textwidth}
        \centering
        \includegraphics[width=\linewidth]{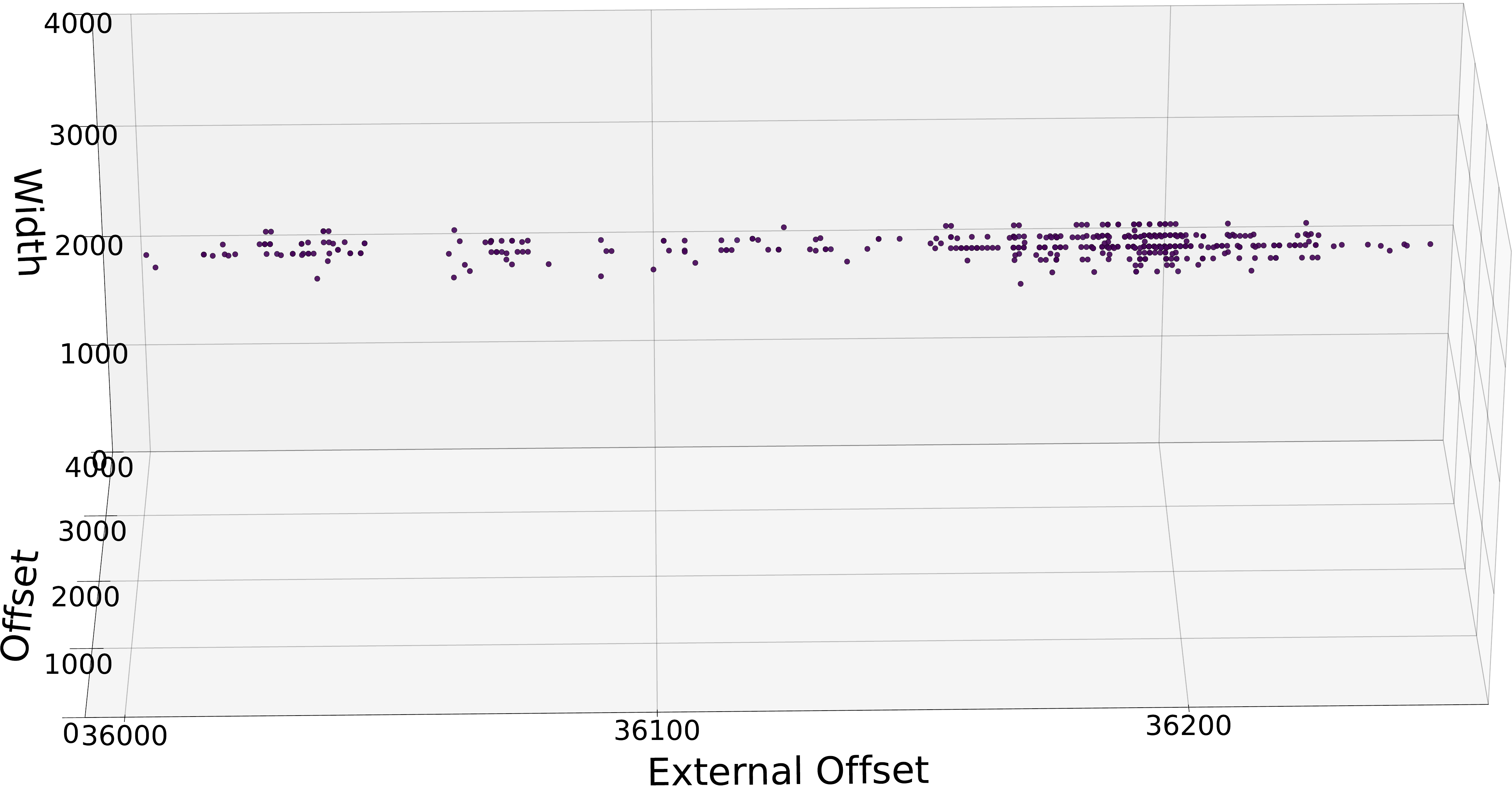}
        \caption{\small Layer 3: Dense 2}
        \label{fig:dense_2_3d}
    \end{subfigure}

    \vspace{0.6em}
    \begin{subfigure}{0.32\textwidth}
        \centering
        \includegraphics[width=\linewidth]{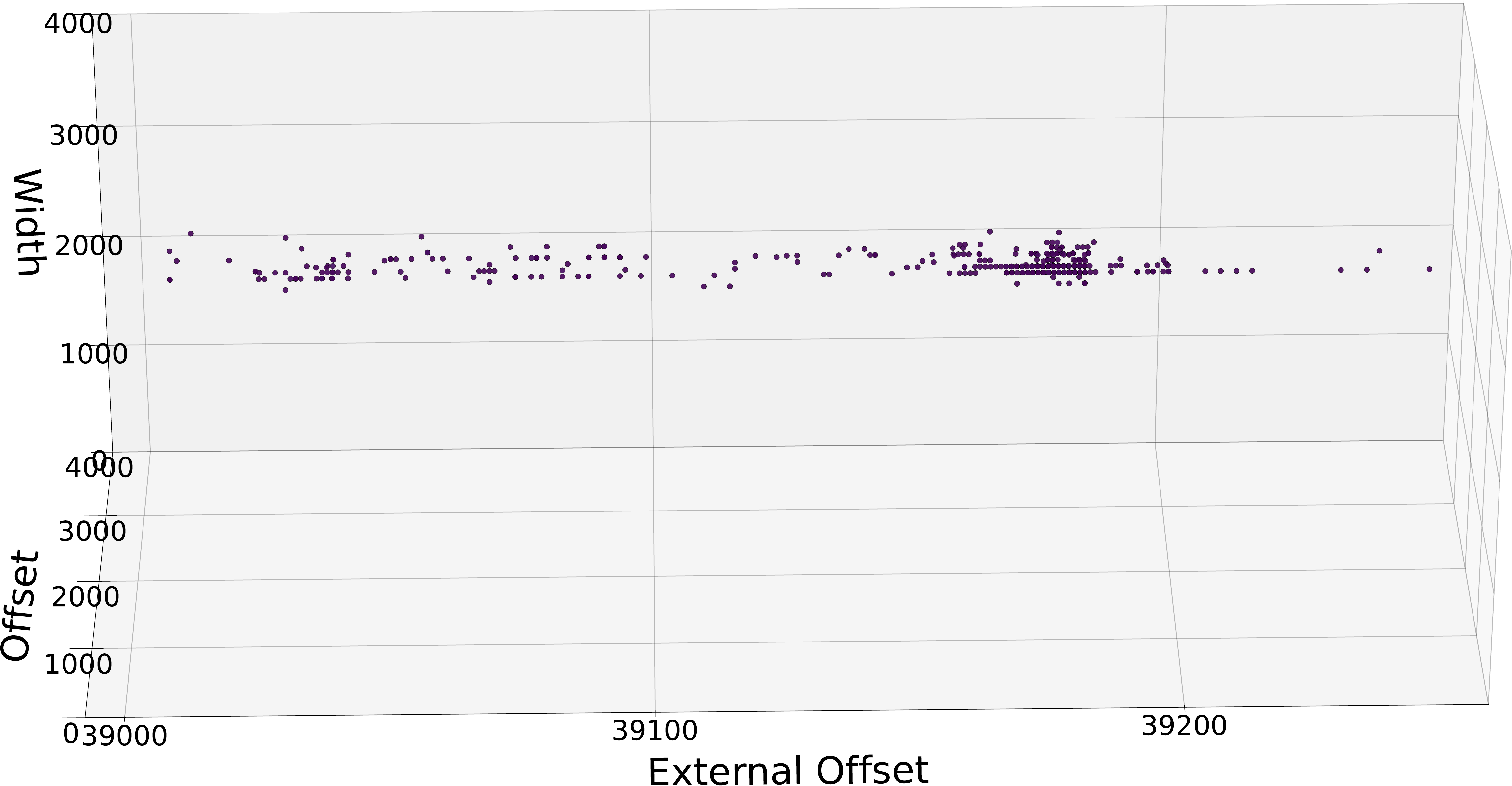}
        \caption{\small Layer 4: ReLU 2}
        \label{fig:relu_2_3d}
    \end{subfigure}\hfil
    \begin{subfigure}{0.32\textwidth}
        \centering
        \includegraphics[width=\linewidth]{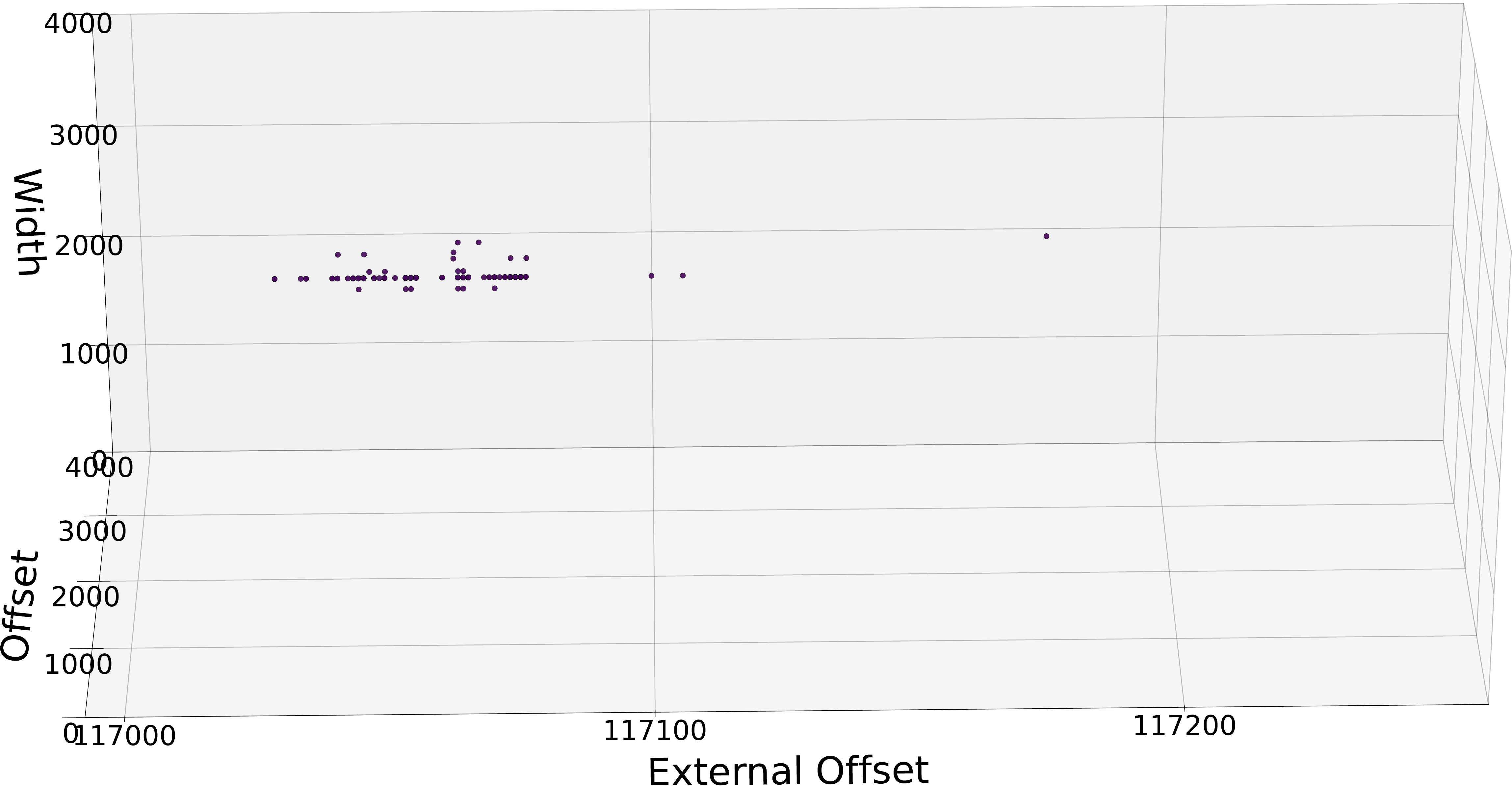}
        \caption{\small Layer 5: Output}
        \label{fig:output_3d}
    \end{subfigure}
    \caption{\small Points where successful voltage glitches were found for each HERQULES layer (axes: \texttt{width}, \texttt{offset}, \texttt{external offset}).}
    \label{fig:scatter_grid}
\end{figure*}

We first present results of finding successful faults across the layers of the neural network in Section~\ref{sec:herqules_fault_setup} responsible for error correction of the quantum computer readout. Figures~\ref{fig:dense_1_3d} to~\ref{fig:output_3d} show the \texttt{width}, \texttt{offset}, and \texttt{external offset} of successful faults; the \texttt{repeat} parameter is omitted from the 3D plots and is reported in Tables~\ref{tab:fault_configs_dense_1} to~\ref{tab:fault_configs_output_1} alongside the top-5 faulting configurations per layer. These experiments focus on the {\em untargeted} faults: reducing the accuracy and increasing the Hamming distance between predicted and true $5$-bit strings, maximally causing the output to change, leading to incorrect interpretation of the quantum computation's results.

We observe that all layers in the HERQULES neural network are susceptible to fault injection. The 3D plots demonstrate that for each layer, the successful faults can be found at variety of external offset. There is some clustering, e.g., {\em Layer 1: Dense 1} and {\em Layer 5: Output} tend to have many faults at offset close to the start of each of the layers (left-side of the graphs), while the other layers tend to have more faults clustered at higher external offset (right-side of the graphs).

Throughout this section we report \emph{fault success} as the number of trials, out of $96$, in which the predicted final output class differs from the ground truth and the device does not reset or hang. The glitch is timed into the cycle window of a specific layer (Algorithm~\ref{alg:herqules_forward_cycles}), but the success count is always measured at the network's final output, so the metric reflects mispredictions that propagated all the way through the remaining layers and softmax rather than transient corruption that was absorbed downstream.

Based on Tables~\ref{tab:fault_configs_dense_1} to~\ref{tab:fault_configs_output_1}, we observe a clear pattern: early layers ({\em Layer 1: Dense 1} and {\em Layer 2: ReLU 1}) yield the highest success rates across the top configurations (up to $27$ of $96$ trials), whereas later layers reach only around $5$ of $96$. Faults timed to earlier layers therefore reach the final output as mispredictions in a large fraction of runs, whereas faults at later stages (e.g., the output layer) are absorbed before reaching the predicted class. This is consistent with the intuition that dense layers execute long MAC chains (large timing surface) and that downstream non-linearities can amplify early numerical perturbations into categorical output collapse~\cite{breier2018faultdnn}.

The results also show that attackers can easily find fault injection points, and many of them result in successful {\em untargeted} attacks. The successful values for \texttt{width} and \texttt{offset} cluster in a narrow range (roughly $2400$ to $2800$ cycles), while the best \texttt{external offset} is layer-dependent (since it indexes into MCU clock cycles from the trigger and so its range is set by where the targeted layer executes within the forward pass).

\begin{table}[t]
  \centering
  \caption{\small Best fault-inducing configuration per layer in the 5-layer network targeting {\em Layer 1: Dense 1} layer.
  The No. of Faults column corresponds to the number of successful faults observed out of \emph{96~repetitions}.}
  \begin{adjustbox}{width=0.48\textwidth}
  \label{tab:fault_configs_dense_1}
  \begin{tabular}{lrrrrr}
    \toprule
    \textbf{Configuration} & \textbf{Width} & \textbf{Offset} & \textbf{External Offset} & \textbf{Repeats} & \textbf{No. of Faults} \\
    \midrule
     1  & 2400 & 2400 & 10026 & 2 & 13 \\
    2   & 2400 & 2400 & 10021 & 2 & 13 \\
    3  & 2400 & 2400 & 10010 & 2 & 12 \\
   4 &  2400 & 2400 & 10013 & 2 & 12 \\
    5  & 2400 & 2500 & 10027 & 2 & 12 \\
    \bottomrule
  \end{tabular}
  \end{adjustbox}
\end{table}

\begin{table}[th!]
  \centering
  \caption{\small Best fault-inducing configuration per layer in the 5-layer network targeting {\em Layer 2: ReLU 1} layer.
  The No. of Faults column corresponds to the number of successful faults observed out of \emph{96~repetitions}.}
  \begin{adjustbox}{width=0.48\textwidth}
  \label{tab:fault_configs_relu_1}
  \begin{tabular}{lrrrrr}
    \toprule
    \textbf{Configuration} & \textbf{Width} & \textbf{Offset} & \textbf{External Offset} & \textbf{Repeats} & \textbf{No. of Faults} \\
    \midrule
     1  & 2700 & 2600 & 14208 & 5 & 27 \\
    2   & 2700 & 2600 & 14207 & 5 & 24\\
    3  & 2700 & 2600 & 14209 & 5 & 23 \\
   4 &  2700 & 2600 & 14206 & 5 & 22 \\
    5  & 2700 & 2600 & 14211 & 5 & 22 \\
    \bottomrule
  \end{tabular}
  \end{adjustbox}
\end{table}

\begin{table}[th!]
  \centering
  \caption{\small Best fault-inducing configuration per layer in the 5-layer network targeting {\em Layer 3: Dense 2} layer.
  The No. of Faults column corresponds to the number of successful faults observed out of \emph{96~repetitions}.}
  \begin{adjustbox}{width=0.48\textwidth}
  \label{tab:fault_configs_dense_2}
  \begin{tabular}{lrrrrr}
    \toprule
    \textbf{Configuration} & \textbf{Width} & \textbf{Offset} & \textbf{External Offset} & \textbf{Repeats} & \textbf{No. of Faults} \\
    \midrule
     1  & 2600 & 2800 & 36170 & 5 & 5 \\
    2   & 2700 & 2600 & 36116 & 5 & 5\\
    3  & 2500 & 2600 & 36194 & 1 & 4 \\
   4 &  2700 & 2600 & 36210 & 4 & 4 \\
    5  & 2600 & 2800 & 36194 & 4 & 4 \\
    \bottomrule
  \end{tabular}
  \end{adjustbox}
\end{table}

\begin{table}[th!]
  \centering
  \caption{\small Best fault-inducing configuration per layer in the 5-layer network targeting {\em Layer 4: ReLU 2} layer.
  The No. of Faults column corresponds to the number of successful faults observed out of \emph{96~repetitions}.}
  \begin{adjustbox}{width=0.48\textwidth}
  \label{tab:fault_configs_relu_2}
  \begin{tabular}{lrrrrr}
    \toprule
    \textbf{Configuration} & \textbf{Width} & \textbf{Offset} & \textbf{External Offset} & \textbf{Repeats} & \textbf{No. of Faults} \\
    \midrule
     1  & 2500 & 2400 & 39175 & 5 & 7 \\
    2   & 2500 & 2400 & 39180 & 5 & 7\\
    3  & 2500 & 2400 & 39174 & 5 & 6 \\
   4 &  2500 & 2500 & 39177 & 5 & 6 \\
    5  & 2500 & 2400 & 39637 & 5 & 6 \\
    \bottomrule
  \end{tabular}
  \end{adjustbox}
\end{table}

\begin{table}[th!]
  \centering
  \caption{\small Best fault-inducing configuration per layer in the 5-layer network targeting {\em Layer 5: Output} layer.
  The No. of Faults column corresponds to the number of successful faults observed out of \emph{96~repetitions}.}
  \begin{adjustbox}{width=0.48\textwidth}
  \label{tab:fault_configs_output_1}
  \begin{tabular}{lrrrrr}
    \toprule
    \textbf{Configuration} & \textbf{Width} & \textbf{Offset} & \textbf{External Offset} & \textbf{Repeats} & \textbf{No. of Faults} \\
    \midrule
     1  & 2500 & 2400 & 117065 & 5 & 4 \\
    2   & 2500 & 2400 & 117044 & 4 & 3\\
    3  & 2500 & 2400 & 117047 & 5 & 3 \\
   4 &  2500 & 2500 & 117050 & 5 & 3 \\
    5  & 2500 & 2400 & 117063 & 5 & 3 \\
    \bottomrule
  \end{tabular}
  \end{adjustbox}
\end{table}

\subsubsection{Fault Impact on Specific Readout Values}

\begin{figure*}[t]
  \centering
    \begin{subfigure}{0.32\textwidth}
        \centering
        \includegraphics[width=\linewidth]{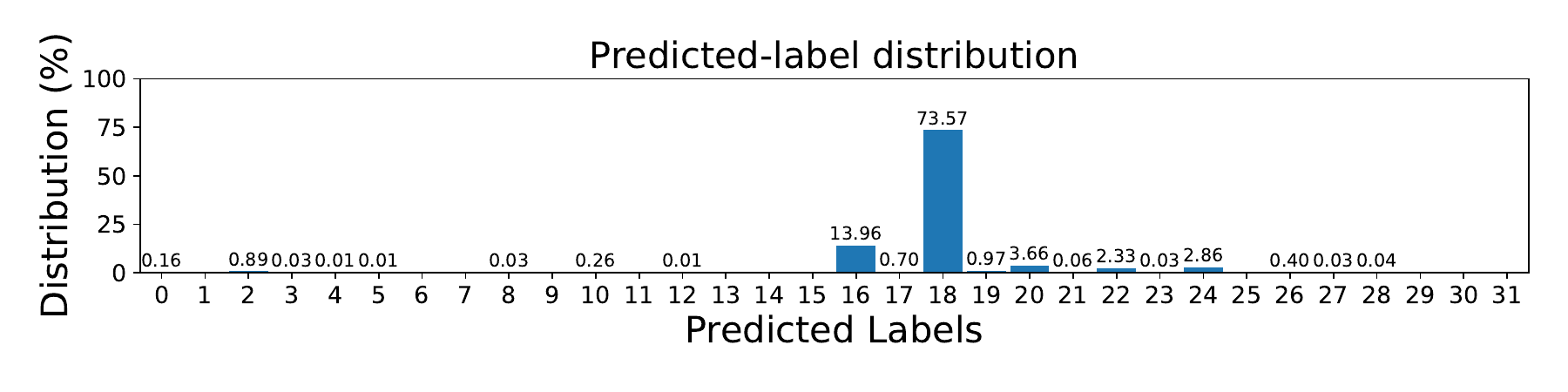}
        \caption{\small No fault}
        \label{fig:18_no_fault}
    \end{subfigure}\hfil
    \begin{subfigure}{0.32\textwidth}
        \centering
        \includegraphics[width=\linewidth]{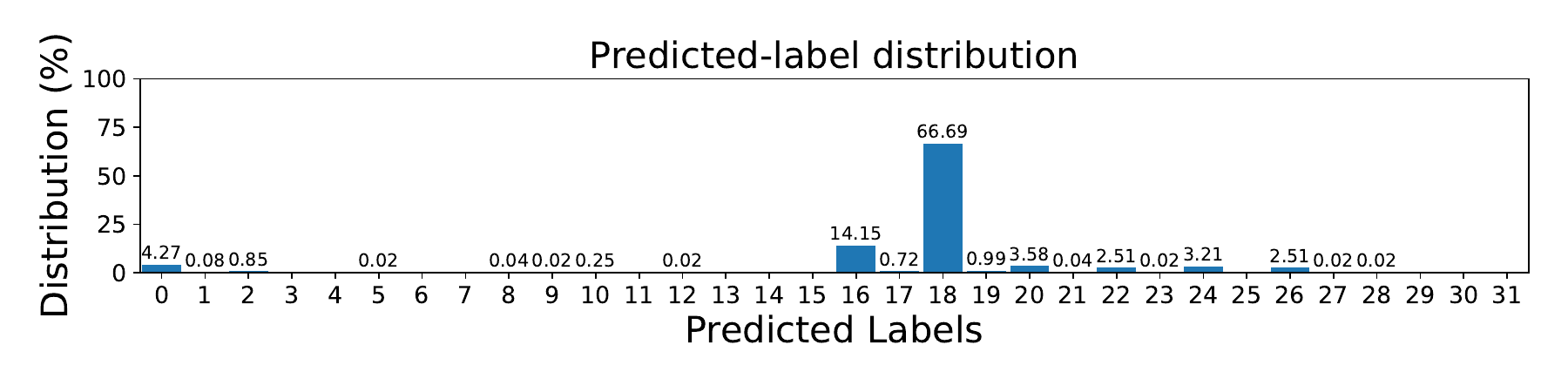}
        \caption{\small Layer 1: Dense 1}
        \label{fig:18_configuration_1_dense_1}
    \end{subfigure}\hfil
    \begin{subfigure}{0.32\textwidth}
        \centering
        \includegraphics[width=\linewidth]{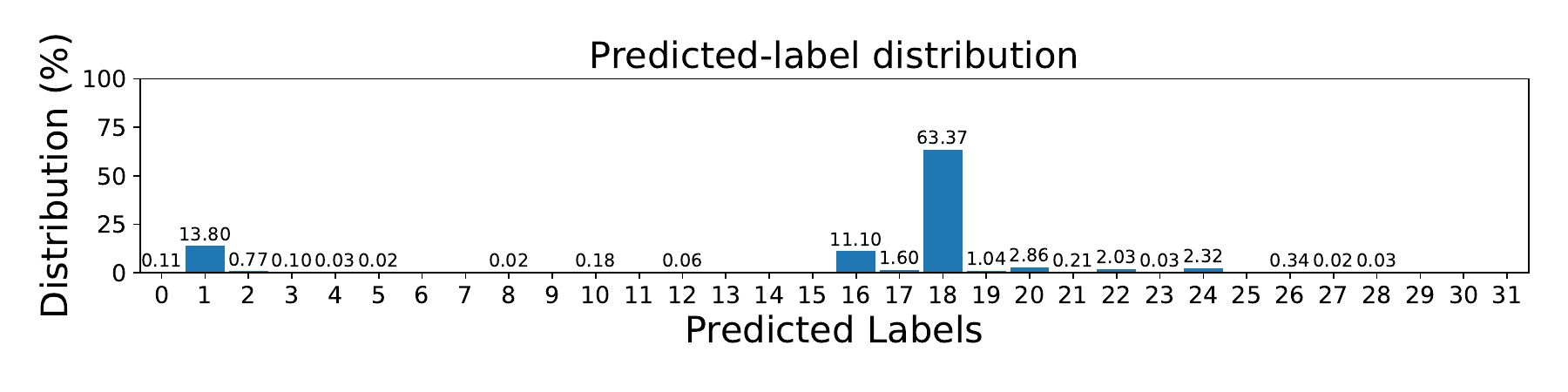}
        \caption{\small Layer 2: ReLU 1}
        \label{fig:18_configuration_1_relu_1}
    \end{subfigure}

    \vspace{0.6em}
    \begin{subfigure}{0.32\textwidth}
        \centering
        \includegraphics[width=\linewidth]{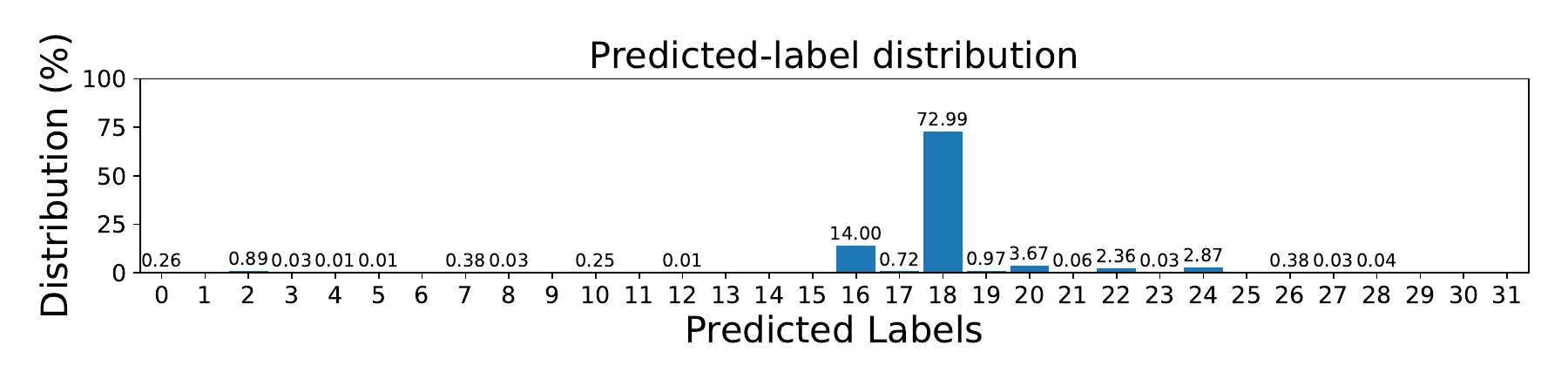}
        \caption{\small Layer 3: Dense 2}
        \label{fig:18_configuration_1_dense_2}
    \end{subfigure}\hfil
    \begin{subfigure}{0.32\textwidth}
        \centering
        \includegraphics[width=\linewidth]{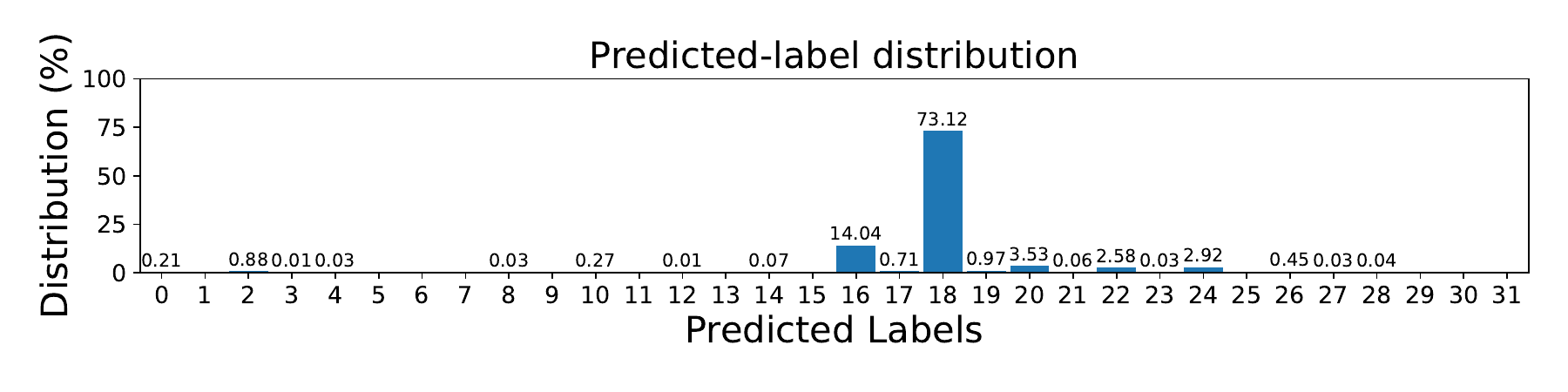}
        \caption{\small Layer 4: ReLU 2}
        \label{fig:18_configuration_1_relu_2}
    \end{subfigure}\hfil
    \begin{subfigure}{0.32\textwidth}
        \centering
        \includegraphics[width=\linewidth]{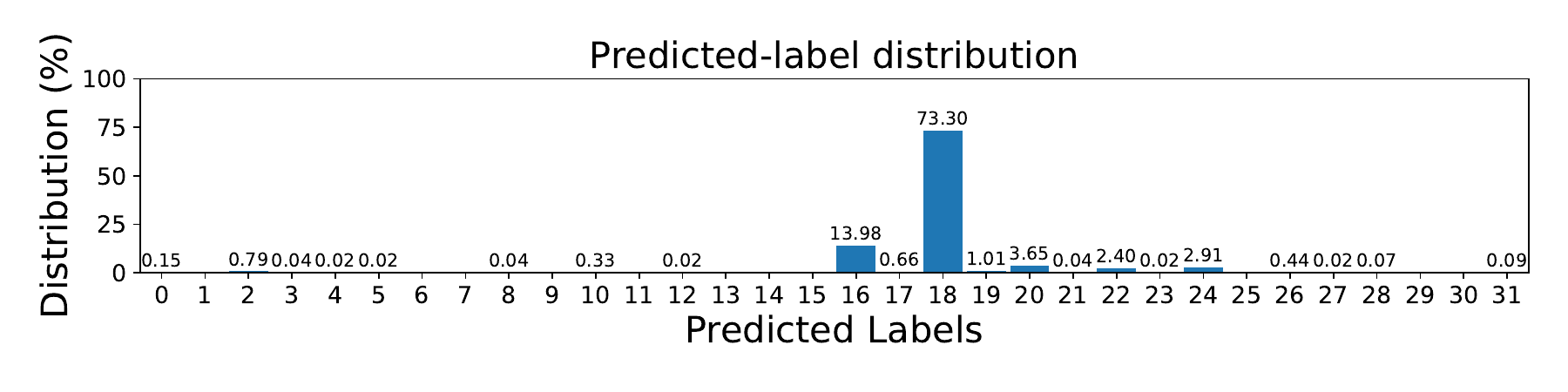}
        \caption{\small Layer 5: Output}
        \label{fig:18_configuration_1_output}
    \end{subfigure}
    \caption{\small Distribution of classification outputs for input $10010$ under no fault and when faulting each HERQULES layer.}
    \label{fig:hist_grid}
\end{figure*}

Beyond untargeted degradation, we examined how the best faulting configurations can bias outputs toward attacker-chosen targets (e.g., forcing specific $5$-bit strings). The results are shown in Figures~\ref{fig:18_no_fault} to~\ref{fig:18_configuration_1_output}.

In the figures we apply the top faulting configuration from each layer onto a randomly selected input from the data set, IQ data whose ground-truth readout is binary $10010$ (integer $18$). Due to a known issue with the device used by HERQULES to collect the data, bit~$1$ is noisy, so even with no fault attack the ML algorithm outputs $18$ with high probability but also frequently outputs $16$ (Figure~\ref{fig:18_no_fault}). Under the {\em Layer 1: Dense 1} fault, output $0$ (binary $00000$) rises from $0.16\%$ at baseline to about $4\%$ (Figure~\ref{fig:18_configuration_1_dense_1}), and under the {\em Layer 2: ReLU 1} fault output $1$ (binary $00001$) appears about $14\%$ of the time (Figure~\ref{fig:18_configuration_1_relu_1}). The correct output $18$ drops from about $74\%$ at baseline to $67\%$ and $63\%$ under the Layer~1 and Layer~2 faults respectively, while remaining around $73\%$ under faults at Layers~3, 4, and 5 (consistent with the prior observation that early layers are most susceptible).

Our results demonstrate that {\em targeted} steering (biasing corrected outputs toward attacker-chosen bitstrings) is harder to achieve than untargeted degradation. Prior work has demonstrated adversarial steering on other ML architectures~\cite{breier2018faultdnn}, but the same fault parameters do not produce equally strong steering on HERQULES when applied to randomly selected inputs. If the attacker knows the input, extended parameter search targeted at that single input could in principle realize targeted steering; we leave this to future work.

\subsection{Deep Q Decoder: Fault Injection Results}
\label{sec:deepq_results}

\begin{figure*}[t]
    \centering
    \begin{subfigure}{0.24\textwidth}
        \includegraphics[width=\textwidth]{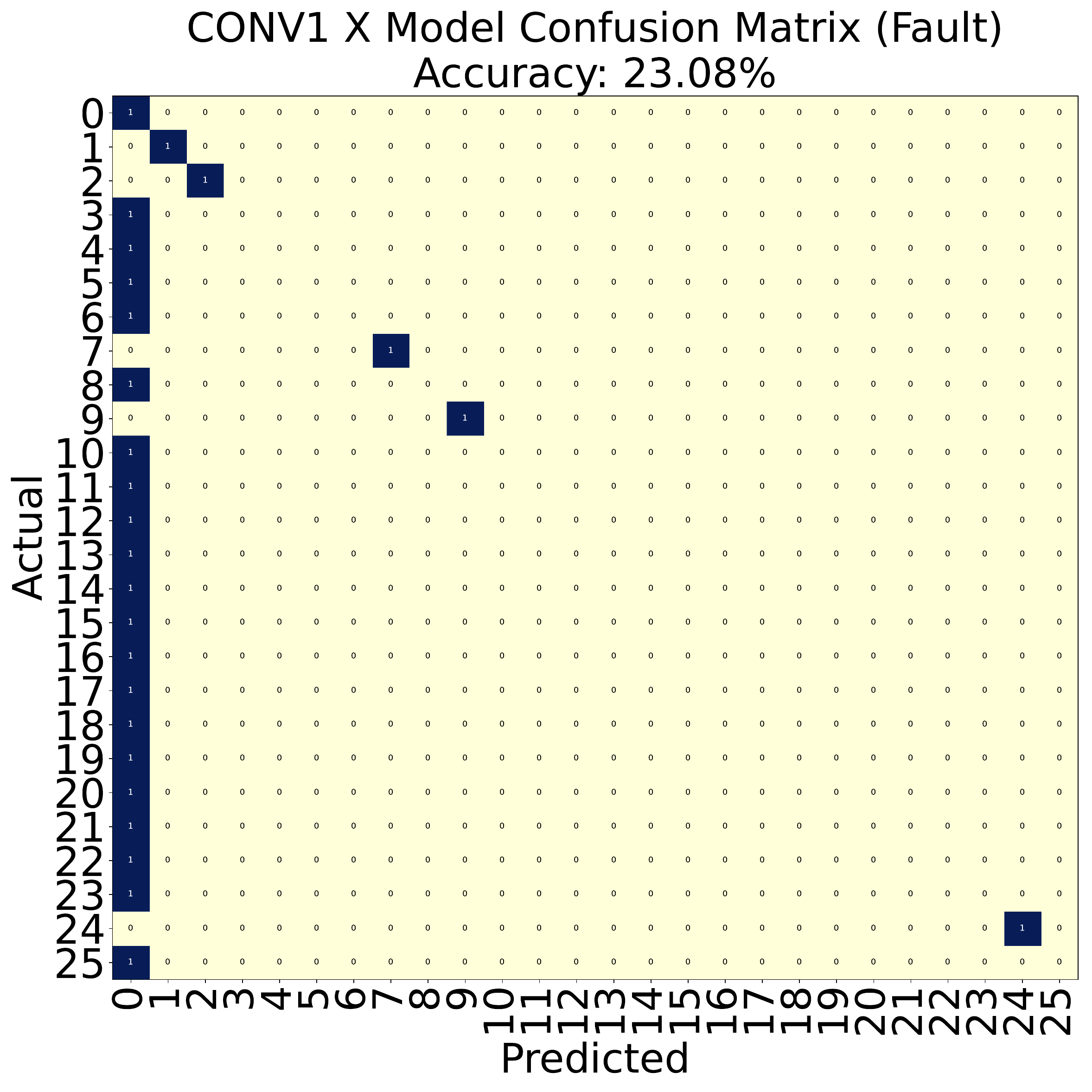}
        \caption{Convolution X Model}
    \label{fig:deepq_conv_x_fault}
    \end{subfigure}
    \hfill
    \begin{subfigure}{0.24\textwidth}
        \includegraphics[width=\textwidth]{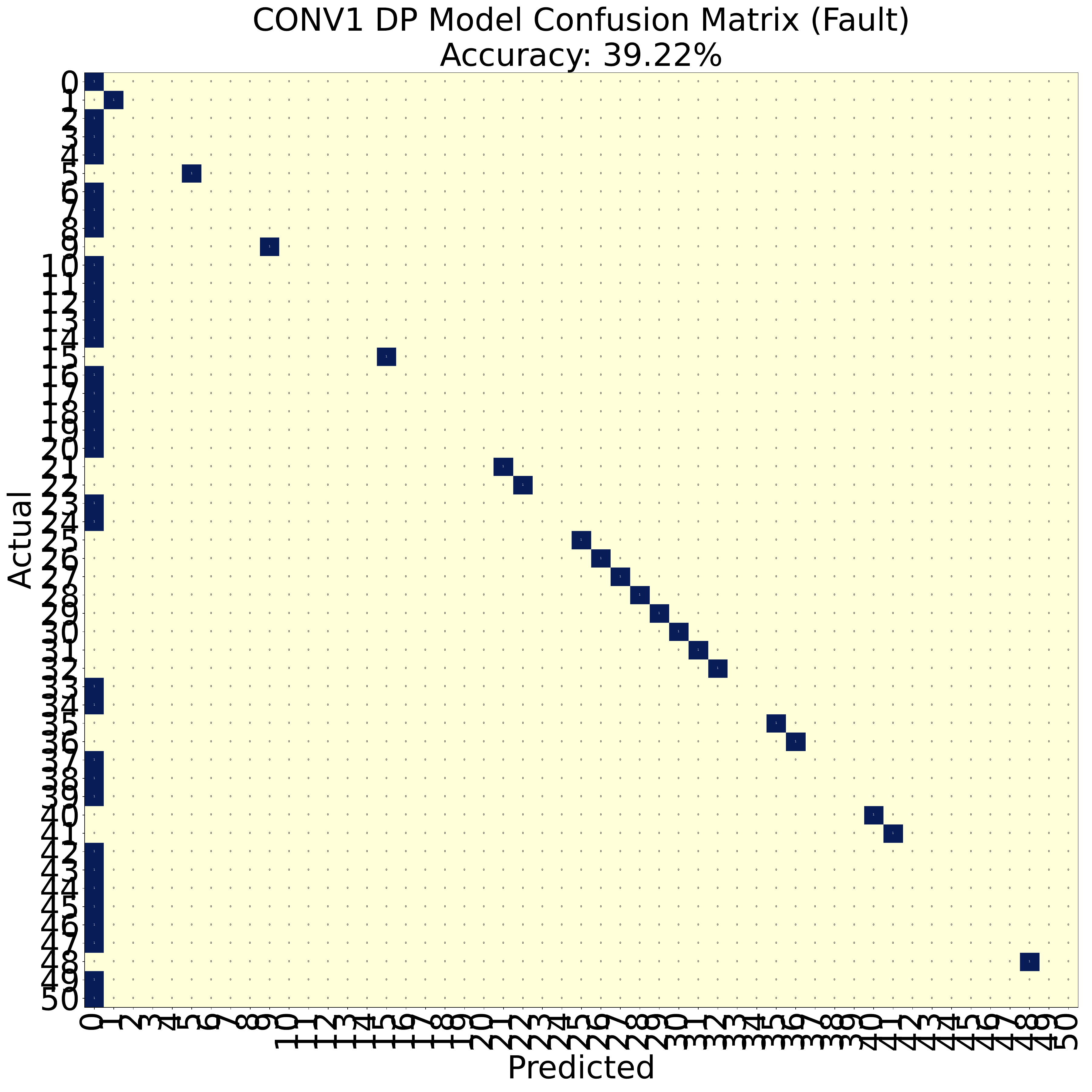}
        \caption{Convolution DP Model}
        \label{fig:deepq_conv_dp_fault}
    \end{subfigure}
    \hfill
    \begin{subfigure}{0.24\textwidth}
        \includegraphics[width=\textwidth]{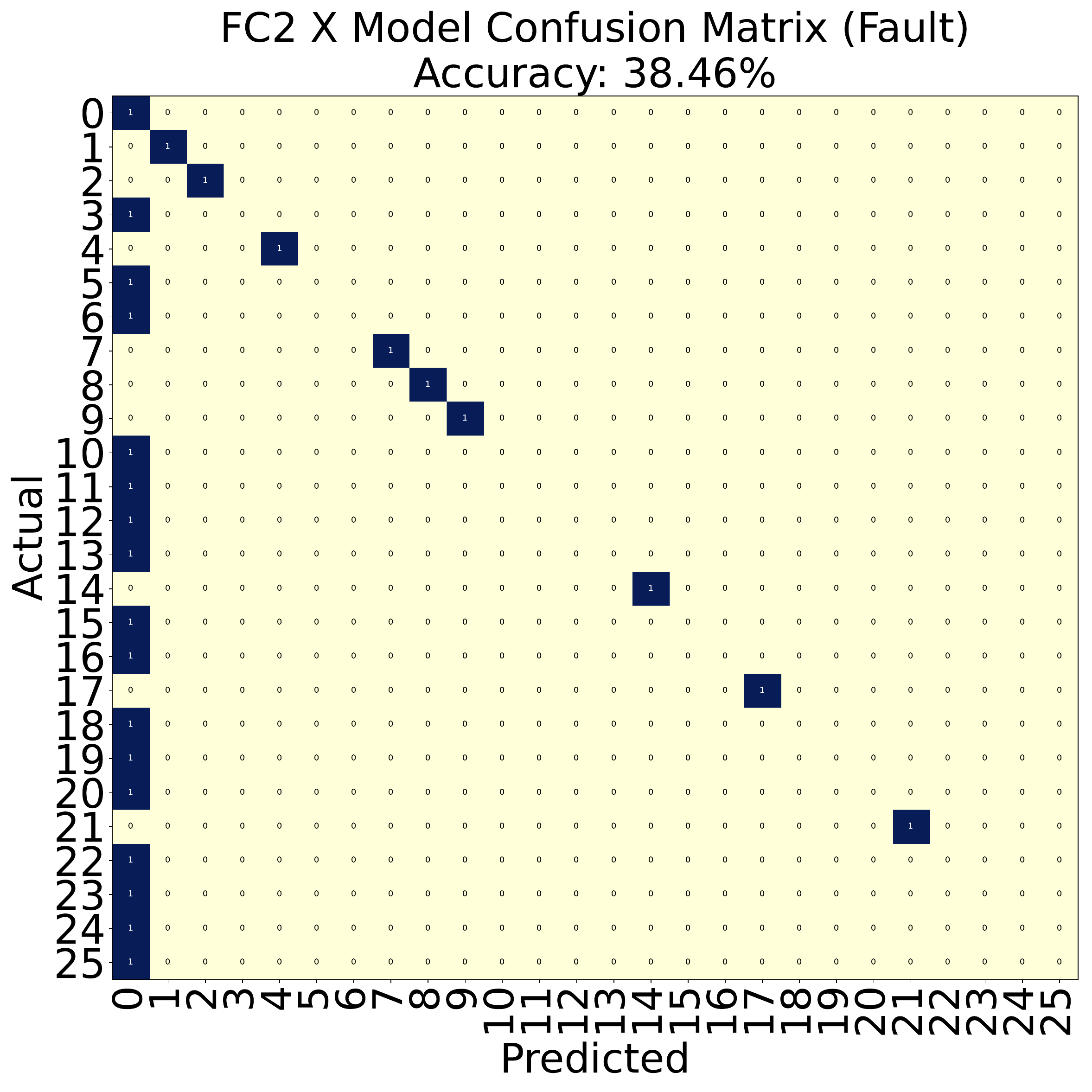}
        \caption{FC2 X Model}
        \label{fig:deepq_fc2_x_fault}
    \end{subfigure}
    \hfill
    \begin{subfigure}{0.24\textwidth}
        \includegraphics[width=\textwidth]{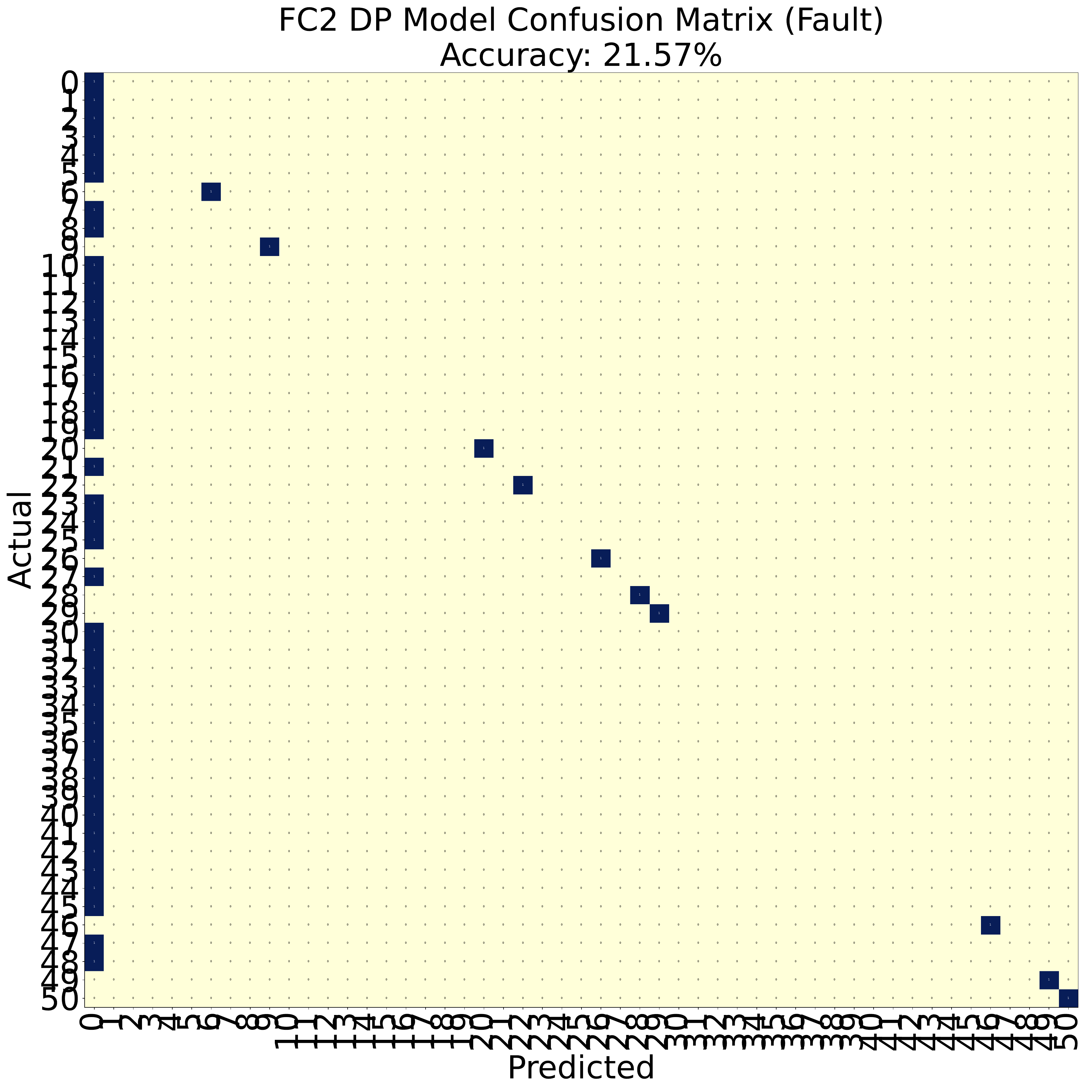}
        \caption{FC2 DP Model}
        \label{fig:deepq_fc2_dp_fault}
    \end{subfigure}
    \caption{Confusion matrices under fault for the Deep~Q decoder, for glitches targeting the first convolutional layer (Conv1) and the dueling output layer (FC2), under both the X-noise and Depolarizing noise models. Each row is the ground-truth corrective action; each column is the action selected under fault. Diagonal cells are correct decodings, off-diagonal cells are fault-induced misdecodings, and a column with concentrated mass indicates the decoder is being steered to that single action regardless of input.}
    \label{fig:deepqdecoding_fault}
\end{figure*}

We now present results from fault-injection experiments on the Deep~Q-learning surface code decoder. As described in Section~\ref{sec:attacksetup}, we target two architecturally distinct layers: the first convolutional layer (Conv1) and the dueling output layer (FC2). We report two sets of results: (i)~hardware fault-injection on Conv1 and FC2, obtained by timing voltage glitches from the ChipWhisperer Husky into the cycle window of the targeted layer, exactly as in the HERQULES study; and (ii)~a complementary software-level non-finite injection at the Conv1 output, in which non-finite values (NaN, $+\infty$, $-\infty$) are written directly into selected positions of the Conv1 output, used to isolate the propagation behavior responsible for the failure mode observed under hardware glitching of Conv1. For both layers, we evaluate two noise-model variants of the trained decoder from~\cite{sweke2021rl_decoders}: an \emph{X-noise} model with $26$ candidate actions, and a \emph{depolarizing} (DP) model with $51$ candidate actions. The confusion matrices presented below treat each row as the ground-truth corrective action that the decoder should select for a given syndrome, and each column as the action actually selected by the embedded implementation; the diagonal therefore corresponds to correct decoding decisions, and any off-diagonal mass is a misdecoding induced by the fault. We separately verified that the C port of each decoder variant reproduces the reference behavior of the trained model with $100\%$ accuracy in the absence of any fault, so all off-diagonal mass in the matrices below is attributable to the injected glitch rather than to residual training error.

\subsubsection{Conv1 Layer Faults}
\label{sec:deepq_conv1}

We evaluate the Conv1 layer of the Deep~Q decoder under two noise-model variants used by Sweke~et~al.~\cite{sweke2021rl_decoders}: the \emph{X-noise} model with $26$ candidate actions and the \emph{depolarizing} (DP) noise model with $51$ candidate actions. Figures~\ref{fig:deepq_conv_x_fault} and~\ref{fig:deepq_conv_dp_fault} show the confusion matrices when a single voltage glitch is injected into the Conv1 cycle window using the trigger-synchronized methodology of Section~\ref{sec:attacksetup}. The accuracy of the X-noise decoder drops from $100\%$ to $23.08\%$, and the accuracy of the depolarizing decoder drops from $100\%$ to $39.22\%$. The structure of the failures is not diffuse: in both variants the off-diagonal mass concentrates almost entirely in the leftmost column of the confusion matrix, meaning the decoder predicts \emph{action~$0$} for the majority of syndrome inputs that previously resolved to other actions. This is operationally severe for surface-code decoding, because the decoder is supposed to map each distinct syndrome to a syndrome-specific corrective action; under a successful Conv1 glitch, the decoder instead applies a single fixed action regardless of the input, which is the same kind of constant-output failure mode observed when ML classifiers undergo mode collapse under fault injection~\cite{breier2018faultdnn}.

To corroborate that this collapse to action~$0$ is a property of the convolutional pipeline rather than an artifact of any particular glitch parameter, we additionally perform a software-level simulation in which we directly write non-finite values (NaN, $+\infty$, or $-\infty$) into the first five flattened output positions of Conv1 and let the rest of the network execute normally. The results, summarized in Table~\ref{tab:conv1_nan_inf_faults}, show that for every non-finite value tested and for both noise models, the decoder's final chosen action is action~$0$ on every input that does not natively map to action~$0$. The X-noise model's accuracy drops to $3.85\%$ ($1$ of $26$ test inputs preserved, the one whose ground-truth label was already action~$0$), and the depolarizing model's accuracy drops to $1.96\%$ ($1$ of $51$). The fact that injecting just five non-finite scalars at the output of Conv1 is sufficient to produce the same qualitative failure mode as the hardware glitch is consistent with $\mathrm{NaN}$ or $\pm\infty$ values propagating through the subsequent ReLU and dense layers and biasing the resulting Q-value vector toward a single argmax, and provides a hypothesis that defenses can target with a simple non-finite check at layer~boundaries.

\begin{table}[t]
\centering
\caption{CONV1 non-finite fault injection results. For each model, injecting NaN, $+\infty$, or $-\infty$ into any of the first five flattened CONV1 output positions caused the final prediction to collapse to action 0.}
\label{tab:conv1_nan_inf_faults}
\begin{adjustbox}{width=0.48\textwidth}
\begin{tabular}{llccccc}
\toprule
\textbf{Model} & \textbf{Fault Value} & \textbf{Fault Positions} &
\textbf{Final Chosen Action} & \textbf{Accuracy} &
\textbf{Wrong / Total} & \textbf{Effective / Total} \\
\midrule
X  & NaN        & 0--4 & 0 & 3.846\% & 25 / 26 & 26 / 26 \\
X  & $+\infty$ & 0--4 & 0 & 3.846\% & 25 / 26 & 26 / 26 \\
X  & $-\infty$ & 0--4 & 0 & 3.846\% & 25 / 26 & 26 / 26 \\
\midrule
DP & NaN        & 0--4 & 0 & 1.961\% & 50 / 51 & 51 / 51 \\
DP & $+\infty$ & 0--4 & 0 & 1.961\% & 50 / 51 & 51 / 51 \\
DP & $-\infty$ & 0--4 & 0 & 1.961\% & 50 / 51 & 51 / 51 \\
\bottomrule
\end{tabular}
\end{adjustbox}
\end{table}

\subsubsection{FC2 Layer Faults}
\label{sec:deepq_fc2}

The FC2 layer is the dueling output layer that directly produces the value and per-action advantage outputs from which the Q-values are formed; an argmax over those Q-values selects the corrective action. A fault in this layer can directly alter the argmax over Q-values, causing the decoder to select an incorrect correction. Unlike faults in earlier layers where non-linearities and subsequent processing may attenuate perturbations, faults in FC2 have a direct, unmediated path to changing the decoder's output action.

Figures~\ref{fig:deepq_fc2_x_fault} and~\ref{fig:deepq_fc2_dp_fault} show the confusion matrices when a single voltage glitch is timed into the FC2 cycle window. The X-noise decoder's accuracy drops from $100\%$ to $38.46\%$, and the depolarizing decoder's accuracy drops from $100\%$ to $21.57\%$. As with Conv1, the misclassifications are not uniformly distributed: the dominant failure mode is again a collapse toward the leftmost column, i.e., the decoder is steered into selecting action~$0$ (and, to a lesser extent, a small number of other low-index actions) for inputs that should produce other corrections.

Two empirical observations distinguish FC2 faults from Conv1 faults. First, the relative susceptibility of FC2 versus Conv1 depends on the noise model: on the X-noise model, FC2 retains higher accuracy under fault than Conv1 ($38.46\%$ vs.\ $23.08\%$), while on the depolarizing model the relationship reverses and FC2 is more degraded than Conv1 ($21.57\%$ vs.\ $39.22\%$). Our experiments do not isolate a single mechanism that explains both directions of this comparison. Second, the result for FC2 differs qualitatively from the HERQULES output layer (Layer~5), where attacker success was rare ($\sim$$5/96$ in Table~\ref{tab:fault_configs_output_1}). We note two structural distinctions between the two final-layer settings that may be relevant: HERQULES produces $32$-class softmax-normalized probabilities, while the Deep~Q FC2 layer produces $26$ or $51$ raw, unnormalized Q-values whose argmax is the chosen action.

These results emphasize that, for ML-based error-correction decoders, both early convolutional layers and the final fully connected layer admit single-shot voltage glitches that produce structured, attacker-favorable failure modes rather than purely random noise.

\section{Discussion}

\subsection{HERQULES and Deep Q Decoder Comparison}

Across the two case studies, several differences between the Deep~Q decoder's CNN architecture and HERQULES' fully connected network are worth noting:
\begin{itemize}

\item \textbf{Weight sharing in convolutional layers:} Conv1 reuses kernel weights across all spatial positions, so a fault in a few partial-sum accumulators can corrupt many output activations at once, whereas fully connected layers do not share parameters; which of Conv1 or FC2 is more fragile depends on the noise model (Section~\ref{sec:results}).

    \item \textbf{Silent contamination by non-finite values:} The non-finite simulation results in Table~\ref{tab:conv1_nan_inf_faults} show that writing $\mathrm{NaN}$ or $\pm\infty$ into five Conv1 output positions is enough to drive every subsequent inference to action~$0$. The surrounding control flow does not flag the non-finite intermediate, so this corruption is silent in the sense that the decoder still returns a syntactically well-formed action selection on every input.
        \item \textbf{Output semantics:} HERQULES outputs class probabilities for $32$ readout classes after a softmax. The Deep~Q decoder, by contrast, outputs raw Q-values for $26$ or $51$ actions, and the chosen action is an unnormalized argmax. This structural distinction may contribute to why the FC2 attack outcome differs from the HERQULES output-layer outcome (Table~\ref{tab:fault_configs_output_1}), although our experiments do not isolate a single mechanism.
    \item \textbf{Downstream impact:} Incorrect decoder actions in the surface code can lead to logical errors that accumulate over time, making even low-rate fault injection potentially damaging to the logical qubit over extended computation~\cite{sweke2021rl_decoders}.
\end{itemize}

\subsection{Domain-General vs.\ Quantum-Specific Findings}
\label{sec:discussion_domain}

It is useful to separate which of our observations are generic to fault injection on neural networks and which are specific to the quantum control setting. Two of our findings confirm, in a new domain, the classical DNN fault-injection literature~\cite{breier2018faultdnn}: (i) early, wide layers with long MAC chains present a larger temporal surface and fault more easily than narrow late layers, and (ii) a successful fault frequently collapses the output onto a single dominant class rather than producing uniform noise. Neither the layer dependence nor the mode-collapse behavior is unique to quantum workloads, and we therefore do not claim them as ML-methodological novelties.

What is specific to this domain is the \emph{consequence} of these failures, which is why the same misprediction is far more serious here than in a conventional ML deployment. First, the Deep~Q collapse to a single action is not merely a wrong label: a decoder whose output is independent of the input syndrome no longer performs the syndrome-conditioned correction that the surface code's fault-tolerance argument presupposes, so a well-formed but syndrome-independent action stream silently invalidates the logical-error guarantee rather than degrading a metric. Second, because decoder actions and corrected readouts feed back into subsequent decoding rounds and into calibration loops, an induced bias can compound into logical errors and corrupt calibration state with effects that persist after the attack ends, a feedback pathway with no analogue in one-shot image classification. These consequences, rather than the raw susceptibility numbers, are the security content new to this work.

\subsection{Security Implications}

Building on the domain-specific consequences above, two implications stand out for deployment. First, the rate at which an attacker can drive the decoder into the syndrome-independent regime is a quantity that any logical-error analysis of the protected qubit would need to account for; quantifying it is left to future work. Second, in cloud settings these risks align with broader concerns that quantum computations may rely on untrusted or compromised classical infrastructure~\cite{trochatos2024dynamic,xu2024quantum,xu2023classification}. Together they reinforce that both ML readout correction and ML-based decoding should be treated as security-critical primitives that a physical attacker can \emph{steer} toward specific outcomes, not merely add noise to.

\subsection{Possible Defenses}
\label{sec:defenses}
We recommend lightweight, deployment-friendly defenses inspired by established fault-attack countermeasures~\cite{barenghi2012fi_survey,moro2014countermeasures}:
(1)~redundant inference (majority voting) on a subset of shots;
(2)~cross-checking against a simpler discriminator (a matched-filter baseline for readout, or MWPM for decoding) and rejecting inconsistent outcomes;
(3)~run-time sanity checks on logits, entropy, and activation ranges, including non-finite ($\mathrm{NaN}$, $\pm\infty$) detection at layer boundaries, which Table~\ref{tab:conv1_nan_inf_faults} shows catches the dominant Conv1 failure mode at low cost;
(4)~hardware monitors (brown-out/glitch detectors, clock monitors) that reset and flag anomalous operation; and
(5)~randomized layer-execution timing (jitter) to hinder precise synchronization. For CNN decoders, checking whether the chosen action is the same low-index action across many consecutive distinct syndromes further flags the attacker-favorable failure mode we observe.

\section{Related Work}
\label{sec:related_work}

Section~\ref{sec:background} reviewed ML for quantum readout and QEC decoding, none of which addresses its security. We focus here on the two most closely related areas.

\subsection{Fault Injection Against ML}

Fault attacks on embedded systems have a long history~\cite{boneh1997faults,barenghi2012fi_survey}, and a substantial body of work has since shown that DNN inference is itself vulnerable to physical fault injection: transient voltage, clock, or electromagnetic faults and targeted weight bit-flips can induce severe accuracy degradation and, in some cases, adversary-chosen misclassification on embedded classifiers, accelerators, and microcontroller-based TinyML models~\cite{breier2018faultdnn,etim2026tinyml}. These works target conventional ML tasks (e.g., image classification) where a misprediction only degrades application-level quality, and are distinct from adversarial examples, which perturb the \emph{inputs} of a correctly executing model; here it is the \emph{computation} that is corrupted. We carry this attack class into the quantum control and readout stack, where (Section~\ref{sec:discussion_domain}) an equivalent misprediction is not a quality-of-service issue but a direct threat to the syndrome-conditioned correction and calibration loops the computation depends on. To our knowledge, no prior work has considered physical fault injection against ML used for quantum readout or error-correction decoding.

\subsection{Security of the Quantum Stack}
Work on quantum security spans the software
layer~\cite{ghosh2023primer,tessma2025recovering}, the control pulses
driving qubits~\cite{trochatos2024dynamic,xu2024jailbreaking}, and
crosstalk between
qubits~\cite{deshpande2023design,Tan2026qubithammer,Tan2025readout}.
Our work was inspired by a proposed
classification~\cite{xu2023classification} of fault-injection attacks
on quantum computers; the closest controller-security work targets
side channels~\cite{xu2023exploration,erata2024quantum} rather than
faults, while prior active attacks inject faults at the qubit level
via reset operations~\cite{mi2022securing,tan2023extending} or
crosstalk~\cite{Tan2026qubithammer}, and vulnerabilities of
fault-tolerant readout and decoding have so far only been
explored analytically~\cite{Trochatos2025ftqc}. We are the first to
enact fault injection on a quantum computer \emph{controller}, as an
empirical, hardware-backed case study against the ML components in its
readout and decoding logic.

\section{Conclusion}

We presented the first voltage-glitch fault-injection study against embedded ML models in the quantum computing stack, covering readout error correction (HERQULES) and surface-code decoding (a Deep~Q-learning CNN). Using trigger-synchronized, layer-localized glitches, we showed that all HERQULES layers are vulnerable, early dense layers most so, and that a single glitch in Conv1 or FC2 of the Deep~Q decoder cuts accuracy from $100\%$ to as low as $21.57\%$, collapsing predictions onto one dominant action; a software-level non-finite injection at Conv1 reproduces this. These results motivate treating ML-based readout correction and decoding as security-critical primitives, and future work on additional layers, code distances, and lightweight defenses such as redundant inference, non-finite checks, and hardware glitch monitors.

\section*{Acknowledgements}

This work was supported in part by National Science Foundation grants \nsf{2245344} and \nsf{2332406}.

\flushcolsend

\bibliographystyle{ieeetr}
\bibliography{bibtex/references}

@inproceedings{deshpande2023design,
  title={Design of quantum computer antivirus},
  author={Deshpande, Sanjay and Xu, Chuanqi and Trochatos, Theodoros and Wang, Hanrui and Erata, Ferhat and Han, Song and Ding, Yongshan and Szefer, Jakub},
  booktitle={2023 IEEE International Symposium on Hardware Oriented Security and Trust (HOST)},
  pages={260--270},
  year={2023},
  organization={IEEE}
}

@inproceedings{tessma2025recovering,
  title={Recovering QSVT Polynomials from Side-Channel Information on Quantum Computers},
  author={Tessma, Kidus and Kukina, Hrvoje and Szefer, Jakub},
  booktitle={2025 IEEE 43rd International Conference on Computer Design (ICCD)},
  pages={342--347},
  year={2025},
  organization={IEEE}
}

@inproceedings{oflynn2014chipwhisperer,
  title={Chipwhisperer: An open-source platform for hardware embedded security research},
  author={O’flynn, Colin and Chen, Zhizhang},
  booktitle={International Workshop on Constructive Side-Channel Analysis and Secure Design},
  pages={243--260},
  year={2014},
  organization={Springer}
}

@inproceedings{akiba2019optuna,
  title={Optuna: A next-generation hyperparameter optimization framework},
  author={Akiba, Takuya and Sano, Shotaro and Yanase, Toshihiko and Ohta, Takeru and Koyama, Masanori},
  booktitle={Proceedings of the 25th ACM SIGKDD international conference on knowledge discovery \& data mining},
  pages={2623--2631},
  year={2019}
}

@inproceedings{maurya2023herqules,
  title={Scaling qubit readout with hardware efficient machine learning architectures},
  author={Maurya, Satvik and Mude, Chaithanya Naik and Oliver, William D and Lienhard, Benjamin and Tannu, Swamit},
  booktitle={Proceedings of the 50th Annual International Symposium on Computer Architecture},
  pages={1--13},
  year={2023}
}

@article{lienhard2022dnnreadout,
  title={Deep-neural-network discrimination of multiplexed superconducting-qubit states},
  author={Lienhard, Benjamin and Veps{\"a}l{\"a}inen, Antti and Govia, Luke CG and Hoffer, Cole R and Qiu, Jack Y and Riste, Diego and Ware, Matthew and Kim, David and Winik, Roni and Melville, Alexander and others},
  journal={Physical Review Applied},
  volume={17},
  number={1},
  pages={014024},
  year={2022},
  publisher={APS}
}

@article{maciejewski2020readout,
  title={Mitigation of readout noise in near-term quantum devices by classical post-processing based on detector tomography},
  author={Maciejewski, Filip B and Zimbor{\'a}s, Zolt{\'a}n and Oszmaniec, Micha{\l}},
  journal={Quantum},
  volume={4},
  pages={257},
  year={2020},
  publisher={Verein zur F{\"o}rderung des Open Access Publizierens in den Quantenwissenschaften}
}

@article{nation2021measurement_mitigation,
  title={Scalable mitigation of measurement errors on quantum computers},
  author={Nation, Paul D and Kang, Hwajung and Sundaresan, Neereja and Gambetta, Jay M},
  journal={PRX Quantum},
  volume={2},
  number={4},
  pages={040326},
  year={2021},
  publisher={APS}
}

@article{kim2021deeplearning_readout,
  title={Quantum readout error mitigation via deep learning},
  author={Kim, Jihye and Oh, Byungdu and Chong, Yonuk and Hwang, Euyheon and Park, Daniel K},
  journal={New Journal of Physics},
  volume={24},
  number={7},
  pages={073009},
  year={2022},
  publisher={IOP Publishing}
}

@inproceedings{trochatos2024dynamic,
  title={Dynamic pulse switching for protection of quantum computation on untrusted clouds},
  author={Trochatos, Theodoros and Deshpande, Sanjay and Xu, Chuanqi and Lu, Yao and Ding, Yongshan and Szefer, Jakub},
  booktitle={2024 IEEE International Symposium on Hardware Oriented Security and Trust (HOST)},
  pages={404--414},
  year={2024},
  organization={IEEE}
}

@inproceedings{breier2018faultdnn,
  title={Practical fault attack on deep neural networks},
  author={Breier, Jakub and Hou, Xiaolu and Jap, Dirmanto and Ma, Lei and Bhasin, Shivam and Liu, Yang},
  booktitle={Proceedings of the 2018 ACM SIGSAC Conference on Computer and Communications Security},
  pages={2204--2206},
  year={2018}
}

@article{barenghi2012fi_survey,
   title={Fault injection attacks on cryptographic devices: Theory, practice, and countermeasures},
  author={Barenghi, Alessandro and Breveglieri, Luca and Koren, Israel and Naccache, David},
  journal={Proceedings of the IEEE},
  volume={100},
  number={11},
  pages={3056--3076},
  year={2012},
  publisher={IEEE}
}

@inproceedings{boneh1997faults,
 title={On the importance of checking cryptographic protocols for faults},
  author={Boneh, Dan and DeMillo, Richard A and Lipton, Richard J},
  booktitle={International conference on the theory and applications of cryptographic techniques},
  pages={37--51},
  year={1997},
  organization={Springer}
}

@inproceedings{moro2014countermeasures,
   title={Experimental evaluation of two software countermeasures against fault attacks},
  author={Moro, Nicolas and Heydemann, Karine and Dehbaoui, Amine and Robisson, Bruno and Encrenaz, Emmanuelle},
  booktitle={2014 IEEE International Symposium on Hardware-Oriented Security and Trust (HOST)},
  pages={112--117},
  year={2014},
  organization={IEEE}
}

@article{eslami2020survey_fi,
  title={A survey on fault injection methods of digital integrated circuits},
  author={Eslami, Mohammad and Ghavami, Behnam and Raji, Mohsen and Mahani, Ali},
  journal={Integration},
  volume={71},
  pages={154--163},
  year={2020},
  publisher={Elsevier}
}

@article{gangolli2022iot_fi_review,
  title={A systematic review of fault injection attacks on iot systems},
  author={Gangolli, Aakash and Mahmoud, Qusay H and Azim, Akramul},
  journal={Electronics},
  volume={11},
  number={13},
  pages={2023},
  year={2022},
  publisher={MDPI}
}

@inproceedings{xu2024quantum,
  title={Quantum Computer Fault Injection Attacks},
  author={Xu, Chuanqi and Erata, Ferhat and Szefer, Jakub},
  booktitle={2024 IEEE International Conference on Quantum Computing and Engineering (QCE)},
  volume={1},
  pages={331--337},
  year={2024},
  organization={IEEE}
}

@article{xu2023classification,
  title={Classification of quantum computer fault injection attacks},
  author={Xu, Chuanqi and Erata, Ferhat and Szefer, Jakub},
  journal={arXiv preprint arXiv:2309.05478},
  year={2023}
}

@article{ghosh2023primer,
  title={A primer on security of quantum computing hardware},
  author={Ghosh, Swaroop and Upadhyay, Suryansh and Saki, Abdullah Ash},
  journal={Proceedings of the IEEE},
  year={2025},
  publisher={IEEE}
}

@article{nachman2020unfolding,
  title={Unfolding quantum computer readout noise},
  author={Nachman, Benjamin and Urbanek, Miroslav and de Jong, Wibe A and Bauer, Christian W},
  journal={npj Quantum Information},
  volume={6},
  number={1},
  pages={84},
  year={2020},
  publisher={Nature Publishing Group UK London}
}

@article{rojkov2022bias,
  title={Bias in error-corrected quantum sensing},
  author={Rojkov, Ivan and Layden, David and Cappellaro, Paola and Home, Jonathan and Reiter, Florentin},
  journal={Physical Review Letters},
  volume={128},
  number={14},
  pages={140503},
  year={2022},
  publisher={APS}
}

@article{heinsoo2018rapid,
  title={Rapid high-fidelity multiplexed readout of superconducting qubits},
  author={Heinsoo, Johannes and Andersen, Christian Kraglund and Remm, Ants and Krinner, Sebastian and Walter, Theodore and Salath{\'e}, Yves and Gasparinetti, Simone and Besse, Jean-Claude and Poto{\v{c}}nik, Anton and Wallraff, Andreas and others},
  journal={Physical Review Applied},
  volume={10},
  number={3},
  pages={034040},
  year={2018},
  publisher={APS}
}

@article{sweke2021rl_decoders,
  title={Reinforcement learning decoders for fault-tolerant quantum computation},
  author={Sweke, Ryan and Kesselring, Markus S. and van Nieuwenburg, Evert P. L. and Eisert, Jens},
  journal={Machine Learning: Science and Technology},
  volume={2},
  number={2},
  pages={025005},
  year={2021},
  doi={10.1088/2632-2153/abc609}
}

@article{fitzek2020deepq_toric,
  title={Deep {Q}-learning decoder for depolarizing noise on the toric code},
  author={Fitzek, David and Eliasson, Mattias and Frisk Kockum, Anton and Granath, Mats},
  journal={Physical Review Research},
  volume={2},
  number={2},
  pages={023230},
  year={2020},
  doi={10.1103/PhysRevResearch.2.023230}
}

@article{andreasson2019toric_rl,
  title={Quantum error correction for the toric code using deep reinforcement learning},
  author={Andreasson, Philip and Johansson, Joel and Liljestrand, Simon and Granath, Mats},
  journal={Quantum},
  volume={3},
  pages={183},
  year={2019},
  doi={10.22331/q-2019-09-02-183}
}

@article{bausch2024alphaqubit,
  title={Learning high-accuracy error decoding for quantum processors},
  author={Bausch, Johannes and Senior, Andrew W. and Heras, Francisco J. H. and Edlich, Thomas and Davies, Alex and Newman, Michael and Jones, Cody and Sheridan, Kevin and Vinyals, Oriol and Bacon, Dave},
  journal={Nature},
  volume={635},
  pages={834--840},
  year={2024},
  doi={10.1038/s41586-024-08148-8}
}

@article{lange2023gnn_decoder,
  title={Data-driven decoding of quantum error correcting codes using graph neural networks},
  author={Lange, Moritz and Barzen, Johanna and Leymann, Frank and Vietz, Daniel},
  journal={Physical Review Research},
  volume={7},
  pages={023181},
  year={2025},
  doi={10.1103/PhysRevResearch.7.023181}
}

@inproceedings{erata2024systematic,
  title={Systematic Use of Random Self-Reducibility in Cryptographic Code against Physical Attacks},
  author={Erata, Ferhat and Chiu, TingHung and Etim, Anthony and Nampally, Srilalith and Raju, Tejas and Ramu, Rajashree and Piskac, Ruzica and Antonopoulos, Timos and Xiong, Wenjie and Szefer, Jakub},
  booktitle={Proceedings of the 43rd IEEE/ACM International Conference on Computer-Aided Design},
  pages={1--9},
  year={2024}
}

@INPROCEEDINGS {etim2026tinyml,
author = { Etim, Anthony and Nampally, Srilalith and Rasouli, Aubtin and Mazza, Dustin and Chilakapati, Krishna and Chiu, Tinghung and Erata, Ferhat and Nazhandali, Leyla and Xiong, Wenjie and Szefer, Jakub },
booktitle = { 2026 IEEE International Symposium on Hardware Oriented Security and Trust (HOST) },
title = {{ Fault Injection Attacks and Countermeasures on TinyML Algorithms }},
year = {2026},
volume = {},
ISSN = {},
pages = {68-78},
doi = {10.1109/HOST68814.2026.11604834},
url = {https://doi.ieeecomputersociety.org/10.1109/HOST68814.2026.11604834},
publisher = {IEEE Computer Society},
address = {Los Alamitos, CA, USA},
month =May}

@inproceedings{xu2023exploration,
  author={Chuanqi Xu and Ferhat Erata and Jakub Szefer},
  title={Exploration of Power Side-Channel Vulnerabilities in Quantum Computer Controllers},
  booktitle={Proceedings of the Conference on Computer and Communications Security},
  year={2023},
  month={November}
}

@inproceedings{erata2024quantum,
  author={Ferhat Erata and Chuanqi Xu and Ruzica Piskac and Jakub Szefer},
  title={Quantum Circuit Reconstruction from Power Side-Channel Attacks on Quantum Computer Controllers},
  booktitle={Transactions on Cryptographic Hardware and Embedded Systems},
  month={September},
  year={2024}
}

@article{xu2024jailbreaking,
  author={Xu, Chuanqi and Szefer, Jakub},
  title={Jailbreaking Quantum Computers},
  journal={arXiv preprint arXiv:2406.05941},
  year={2024}
}

@inproceedings{Tan2026qubithammer,
  author={Tan, Yizhuo and Choudhury, Navnil and Basu, Kanad and Szefer, Jakub},
  title={{QubitHammer}: Remotely Inducing Qubit State Change on Superconducting Quantum Computers},
  booktitle={Proceedings of the IEEE International Symposium on Hardware Oriented Security and Trust (HOST)},
  year={2026},
  month={May}
}

@article{tan2023extending,
  title={Extending and Defending Attacks on Reset Operations in Quantum Computers},
  author={Tan, Jerry and Xu, Chuanqi and Trochatos, Theodoros and Szefer, Jakub},
  journal={arXiv preprint arXiv:2309.06281},
  year={2023}
}

@inproceedings{mi2022securing,
  title={Securing NISQ Quantum Computer Reset Operations Against Higher Energy State Attacks},
  author={Chuanqi Xu and Jessie Chen and Allen Mi and Jakub Szefer},
  booktitle={Proceedings of the Conference on Computer and Communications Security},
  month={November},
  year={2023}
}

@inproceedings{Trochatos2025ftqc,
  author={Trochatos, Theodoros and Kang, Christopher and Chong, Frederic T. and Szefer, Jakub},
  title={Exploration of Vulnerabilities of Fault-Tolerant Quantum Computing},
  booktitle={Proceedings of the 26th International Symposium on Quality Electronic Design (ISQED)},
  year={2025},
  month={April}
}

@inproceedings{Tan2025readout,
  author={Tan, Yizhuo and Szefer, Jakub},
  title={{I Know What You Are Reading}: Evaluating Readout Crosstalk in Cloud-based Quantum Computers},
  booktitle={Proceedings of the 2025 Quantum Security and Privacy Workshop (QSec '25)},
  year={2025},
  month={October},
  pages={1--6}
}

\end{document}